\newcommand{\system}{\textsc{Tempo}}

\documentclass[sigconf, screen]{acmart}
\AtBeginDocument{%
  }

\usepackage{forest}
\usepackage{enumitem}
\usepackage{xcolor}

\copyrightyear{2026}
\acmYear{2026}
\setcopyright{cc}
\setcctype{by}
\acmConference[UIST '26]{The 39th Annual ACM Symposium on User Interface Software and Technology}{November 02--05, 2026}{Detroit, MI, USA}
\acmBooktitle{The 39th Annual ACM Symposium on User Interface Software and Technology (UIST '26), November 02--05, 2026, Detroit, MI, USA}
\acmDOI{10.1145/3830398.3830630}
\acmISBN{979-8-4007-2856-3/2026/11}

\begin{document}

\title{``What Are You Really Trying to Do?'': Co-Creating Life Goals from Everyday Computer Use}
\begin{abstract}
Recent advances in user modeling make it feasible to conduct open-ended inference over a person's everyday computer use. Despite longstanding visions of systems that deeply understand our actions and the purposes they serve in our lives, existing systems only capture what a person is doing in the moment, not why they are doing it, limiting these systems to surface-level support. We introduce striving co-creation, a process for inferring broader life goals from unstructured observations of computer use. Grounded in Activity Theory and Emmons' personal strivings framework, our system progressively constructs a hierarchical representation of a person's activities. Strivings are, however, difficult to fully resolve from observation alone, as the same action can be driven by many different goals. Our system therefore supports an editing interface that gives people agency over how they are understood by the system, feeding their corrections back into subsequent rounds of striving induction. In a week-long field deployment ($N=14$), we find that our co-creation process produces strivings that participants recognize as representative of their long-term goals and gives them greater agency than baseline methods.
\end{abstract}

\author{Shardul Sapkota}
\affiliation{%
  \institution{Stanford University}
  \city{Stanford}
  \country{USA}
}
\email{sapkota@stanford.edu}

\author{Matthew J{\"o}rke}
\authornote{Both authors contributed equally to this research.}
\affiliation{%
  \institution{Stanford University}
  \city{Stanford}
  \country{USA}
}
\email{joerke@stanford.edu}

\author{Zane Sabbagh}
\authornotemark[1]
\affiliation{%
  \institution{Stanford University}
  \city{Stanford}
  \country{USA}
}
\email{zanesabbagh@stanford.edu}

\author{Omar Shaikh}
\affiliation{%
  \institution{Stanford University}
  \city{Stanford}
  \country{USA}
}
\email{oshaikh@stanford.edu}

\author{Grace Wang}
\affiliation{%
  \institution{Stanford University}
  \city{Stanford}
  \country{USA}
}
\email{gracewng@stanford.edu}

\author{James A. Landay}
\affiliation{%
  \institution{Stanford University}
  \city{Stanford}
  \country{USA}
}
\email{landay@stanford.edu}
\renewcommand{\shortauthors}{Sapkota et al.}


% \section{CCS Concepts and User-Defined Keywords}

\begin{CCSXML}
<ccs2012>
   <concept>
       <concept_id>10003120.10003121.10003129</concept_id>
       <concept_desc>Human-centered computing~Interactive systems and tools</concept_desc>
       <concept_significance>500</concept_significance>
       </concept>
 </ccs2012>
\end{CCSXML}

\ccsdesc[500]{Human-centered computing~Interactive systems and tools}

\keywords{User Models, Context-Aware Computing, Goal Elicitation}
\begin{teaserfigure}
  \centering
  \includegraphics[width=\textwidth]{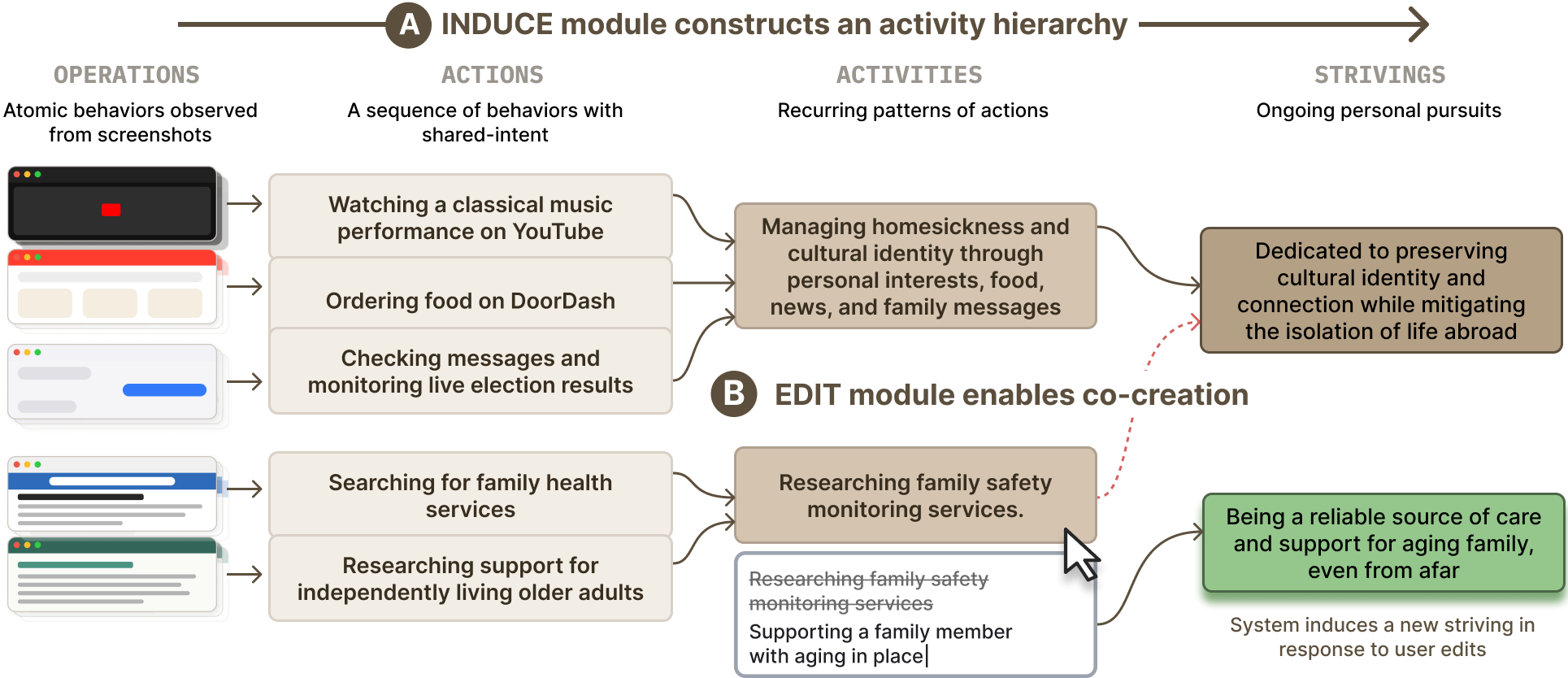}
  \caption{Striving co-creation is a process in which a person and a system jointly construct a representation of the person's long-term life goals from unstructured computer use. We instantiate this process in \system{}, a system that consists of two modules: (A) An Induce module, which progressively abstracts passive screenshots into a four-level hierarchy, and (B) an Edit module, which lets users refine the hierarchy and feeds their corrections into future inductions.}
  \label{fig:teaser}
\end{teaserfigure}

\maketitle
\section{Introduction}

How might we build systems that understand a person's immediate actions in relation to their broader life goals? Consider, for example, someone researching a city they have never visited. They might update a spreadsheet to see whether they can afford the move and text their partner about local schools. While a system observing these interactions would notice fragmented actions across different apps and timeframes, these actions represent a single activity of planning a relocation. That activity may, in turn, support a broader personal pursuit, such as trying to build a stable home for a growing family. Human activity draws much of its meaning from exactly this kind of hierarchical organization, in which low-level actions serve higher-level activities, and activities serve broader, long-term strivings. From Weiser's ubiquitous computing~\cite{weiser1999ubicomp} to Apple's Knowledge Navigator~\cite{apple1987knowledge}, the most enduring visions of human-computer interaction have imagined systems that understand the activities that structure everyday life. Such an understanding of how immediate actions relate to long-term life goals could enable proactive agents~\cite{maes1994agents}, reflection tools~\cite{li2010stage, epstein2015lived}, adaptive interfaces~\cite{gajos2004supple}, or other context-aware systems~\cite{dey2001understanding, abowd1999towards} that help us make sense of what matters and pursue our long-term goals.

\looseness=-1 Today's applications fall short of this vision because their representations of context capture only the lower levels of the hierarchy. The dominant signals available to these systems (e.g., location, device state, application focus, and in-app behavior) describe what a person is doing, but not \textit{why}. For someone planning a relocation, a system limited to such surface-level understanding might suggest better apartment filters or even generate a new interface on the fly to complete a task faster. Yet it would not know to surface school districts suited to the family's needs or connect the move to an ongoing effort to reduce financial stress. While recent progress in user modeling has enabled systems to reason across applications and infer a user's tasks and state from observed activity~\cite{shaikh2025creating, lam2025just, zhang2025summact}, these inferences remain largely task-level and situational. More broadly, computational approaches to goal inference focus on immediate actions~\cite{zhang2025summact, lam2025just, coscia2025ongoal, lu2024proactive}, ask users to articulate their own goals upfront~\cite{choi2025state, jones2024designing}, or hand-define activity structures~\cite{kaptelinin2003umea, bardram2009activity} rather than inferring the hierarchy from observed behavior. What is missing is a representation that synthesizes behavior across domain and time. Such a representation could recognize when one pursuit conflicts with another or when scattered activities serve a single motivation. It could also notice gradual neglect of a pursuit through the behavior hierarchy rather than any single recurring task.  

Still, hierarchical inference alone is insufficient because the mapping from behavior to goals is often \emph{ambiguous}. The person in our example could be building stability for a growing family or leaving a community where they no longer belong. More data or better models cannot always resolve this ambiguity. If the system infers goals on its own, it must pick one plausible interpretation and impose that choice on the user. Yet restricting the system to user-articulated goals gives up inferences the person has not yet put into words, or even recognized. Instead, we argue that goals must be \textbf{\textit{co-created}} with the user. On a technical level, user input addresses both the cold-start problem and the disambiguation problem. More importantly, co-creation ensures that the person, not the system, has the final say over how they are represented and understood by a system that takes action on their behalf. 

In this paper, we introduce \textbf{\textit{striving co-creation}}, a computational process in which a person and the system jointly construct, from unstructured computer use, a representation of the long-term goals that person is working toward. The system synthesizes actions across applications and time into higher-order goal statements, and the person reshapes those statements to reflect their own understanding. For instance, a system observing the spreadsheet, school-district searches, and partner messages from our earlier example might initially synthesize the actions as \texttt{trying to relocate to a more affordable city}. The person might recognize that the move serves a broader purpose and reshape it as \texttt{trying to build a stable home for a growing family}. The system carries that edit forward as a constraint on subsequent inductions.

To instantiate striving co-creation, we develop a system called \system{}. Drawing on Activity Theory's account of hierarchical action~\cite{leontiev1978activity, kaptelinin2009acting} and Emmons' concept of personal strivings~\cite{emmons1986personal}, \system{} organizes observed behavior into a hierarchy of operations, actions, activities, and strivings. Our system consists of two main modules. First, an \textit{Induce} module begins with a self-description provided by the user and progressively abstracts screen observations into this hierarchy. Then, an \textit{Edit} module surfaces this inferred hierarchy to the person and feeds their corrections back into subsequent induction as constraints. Each synthesis run reflects both what the system observed and what the person knows about themselves. \system{} thus treats strivings as interpretations to be negotiated with rather than objective facts. We also include a contextual integrity~\cite{nissenbaum2004privacy} \textit{Audit} module, inspired by prior work~\cite{shaikh2025creating}, which filters observations the user would not expect to be incorporated into the model.

We evaluate \system{} through a week-long field deployment with 14 participants. We assess the quality of the strivings produced by the system and the utility of the hierarchical structure for negotiating them through editing. Despite the difficulty of inferring long-term goals from noisy screenshots of everyday computer use, participants judged strivings generated by the full system as accurate ($M=1.25$ on a $-3$ to $+3$ scale; 80\% rated $\geq 1$). We find that a week of observation can surface pursuits that extend beyond the observation window, though validating their stability over months requires a longer deployment. Our ablations further show that incorporating user-provided context improves individual striving quality while hierarchical structure improves the representativeness of the striving set. Participants rated the editing experience significantly higher when strivings were represented hierarchically. These findings suggest that modeling what people are working toward over the long term requires both hierarchical inference and co-creative mechanisms that let users shape how they are represented. 
\section{Related Work}

In this section, we situate our work within prior literature on user modeling, human-AI interaction, and reflection on personal data.

\subsection{User Modeling from Behavioral Traces}

We are motivated by a longstanding agenda in HCI to build computational representations of users from observation rather than from explicit specification alone. Our work builds on a long line of work in context-aware and ubiquitous computing, which focuses on sensing user state from behavior signals and adapting system behavior accordingly~\cite{horvitz2013lumiere,maes1994agents,dey2001understanding,abowd1999towards}. These systems can now gather cross-application behavioral observations from interaction logs~\cite{r2024entity, reeves2021screenomics}, while advances in multimodal foundation models have enabled user modeling from unstructured observations such as screenshots~\cite{shaikh2025creating, littlebird2026}. While these systems build increasingly rich representations of users, the representations only capture what a user is doing or interested in without connecting observations to their long-term pursuits. To create meaningful representations of such signals, our work draws from activity-centric computing, which operationalizes Activity Theory to organize work and interactions around higher-level activity structures rather than individual applications~\cite{bardram2009activity,kaptelinin2003umea,li2008activity}. 

A related line of work induces user goals from behavioral traces by matching observed actions to pre-specified libraries~\cite{kautz1986generalized} or  constructing probabilistic models of user intent within individual applications~\cite{horvitz2013lumiere}. Subsequent systems have learned to predict user tasks from desktop interaction logs across applications~\cite{dragunov2005tasktracer}, while recent work has used LLMs to move beyond closed goal sets to infer open-ended goals from action sequences and anticipate user needs from observation~\cite{zhang2025summact,lam2025just,lu2024proactive} or from extended conversational interactions~\cite{coscia2025ongoal}. Other works support long-term or open-ended goals by requiring users to articulate them upfront~\cite{choi2025state, jones2024designing}. 

\looseness=-1 In both lines of work, rather than capturing the long-term cross-domain pursuits that characterize what a person is trying to accomplish in life, the goals remain narrowly scoped to a task or a session, or dependent on explicit user specification. To move beyond task-scoped inference, we draw on personal strivings from personality psychology~\cite{emmons1986personal} and the hierarchical decomposition of human activity from Activity Theory~\cite{kaptelinin2009acting,gay2004activity,leontiev1978activity}. Our work extends this hierarchy one level further, connecting observed screen operations, via actions and activities, to the strivings that organize them.

\subsection{Human-AI Interaction}

From interactive machine learning techniques that let users correct model behavior through examples and explanatory feedback~\cite{kulesza2015principles,amershi2014power,smith2020no} to scrutable user models that make system inferences inspectable~\cite{kay2013creating,bull2007student,balog2019transparent}, a range of approaches have sought to give users visibility into and control over AI representations of themselves. However, research on algorithmic awareness has shown that visibility alone is insufficient, as users who discover how systems represent them want to reshape those representations, not just accept or reject the output~\cite{eslami2015always, karizat2021algorithmic}. In fact, contestability frameworks argue that users need mechanisms to challenge the goals, values, and conceptual relationships encoded in systems~\cite{vaccaro2019contestability,hirsch2017designing, lyons2021conceptualising}, especially as models' learned conceptual structures can systematically diverge from human-encoded ones~\cite{boggust2025abstraction}. 

\looseness=-1 Despite this need for visibility and contestability, users today can correct what a system infers about them, and in some cases control whose judgments the system reflects~\cite{gordon2022jury}, but cannot reshape how the system organizes those inferences into a coherent interpretation of who the user is and what they are trying to accomplish. This concern is underscored by growing evidence that AI representations of users' values and identity can shape how users see themselves, even when those representations are incomplete or sycophantic~\cite{yun2026ai, huang2025values, sharma2026s, jakesch2023co}.

\looseness=-1 Recent work has addressed this gap by making AI intermediate representations editable at multiple levels, framing memory updates as conflict resolution, transforming reasoning chains into navigable trees, and organizing LLM outputs into hierarchical information spaces~\cite{vaithilingam2025semantic,pang2025interactive,suh2023sensecape, feng2024cocoa}. We draw from these interactive editing mechanisms to build \system{}, where edits at each level of the hierarchy propagate to future inductions as new observations arrive.

\subsection{Reflection on Personal Data}
\label{sec:rw_reflection}
 
\looseness=-1 Our work draws on the personal informatics literature, which studies how people collect and reflect on personal data~\cite{li2010stage,epstein2015lived}. Across these studies, a persistent finding is that while systems make it easy to collect data, the burden of connecting that data to the goals people actually care about still falls on the user~\cite{cho2022reflection,bentvelzen2022revisiting,fleck2010reflecting,sellen2010beyond,epstein2020mapping}. One line of recent work addresses this gap through LLM-assisted sensemaking, where the systems translate multi-modal sensing streams into higher-level behavioral insights~\cite{li2025vital,choube2025gloss, jorke2025bloom}. Another line of work supports goal-directed sensemaking by letting users define wellbeing goals and teach the system what those goals mean in terms of their personal data~\cite{jorke2023pearl}. Our work extends this literature by inferring a user's long-term goals and grounding reflection in the discrepancy between the system's inferences and the user's self-perception.

\looseness=-1 Drawing on a goal-setting tradition that supports specific, measurable targets~\cite{locke2019development}, a related thread of personal informatics research has made goals central to self-tracking~\cite{schroeder2019examining,niess2018supporting,sefidgar2024migrainetracker,ekhtiar2023goals}. As a result, tracked goals are typically self-reported and health-centered rather than being open-ended, long-term pursuits~\cite{zhu2025systematic}. Our work inverts the relationship between data and goals. Rather than requiring users to specify goals in advance and track data against them, \system{} begins with observed behavior and proposes candidate strivings while allowing users to reshape the result through co-creation.
\section{Tempo: A System for Striving Co-Creation}
\label{sec:system}

We introduce \system{}, a system that operationalizes striving co-creation. \system{} consists of two modules: an \textit{Induce} module that takes screenshots and lightweight user context as input and progressively abstracts raw observations into a four-level hierarchy; and an \textit{Edit} module that surfaces the inferred hierarchy to the user and feeds user corrections back into subsequent rounds of induction. Figure~\ref{fig:teaser} shows a snippet of the striving hierarchy produced by \system{}. Our code is available at \url{https://github.com/StanfordHCI/striving-cocreation}.

\subsection{The Induce Module}
\label{sec:induce}

\begin{figure*}
    \centering
    \includegraphics[width=\textwidth]{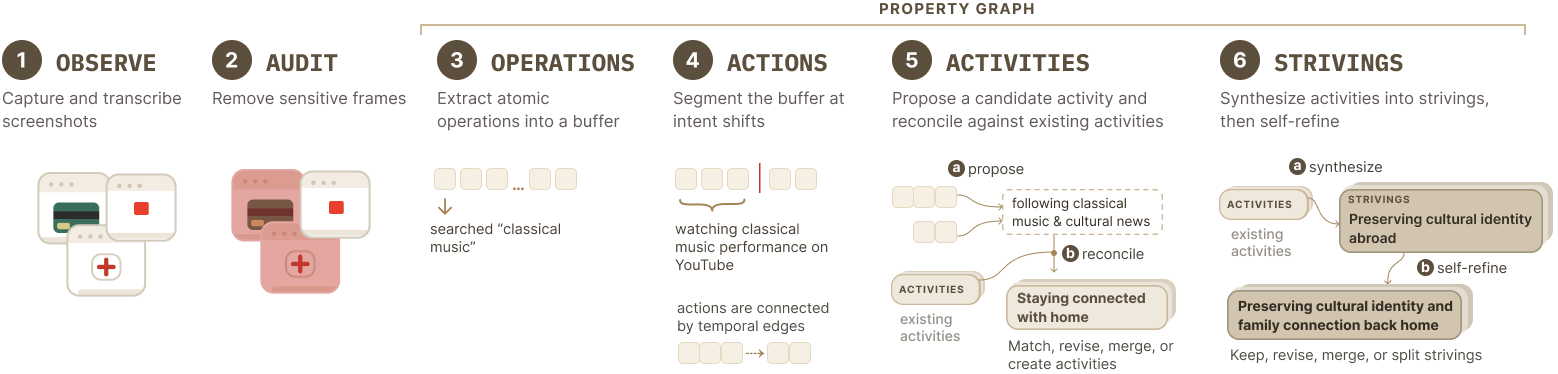}
    \caption{\system{} transcribes screenshots and discards any frame with PII or content blocked by its contextual integrity audit (\S\ref{sec:system_auto_privacy}). It abstracts retained observations into operations, actions, activities, and strivings in a property graph. Activities are proposed and reconciled while strivings are synthesized and self-refined as the model critiques its own output (\S\ref{sec:induce}).}
    \label{fig:induce_module}
\end{figure*}

\system{}'s Induce module progressively abstracts raw screen observations into a four-level hierarchy of behavior. We use Gemini 3 Flash~\cite{team2024gemini} for both screenshot transcription and hierarchical decomposition (see~\S\ref{sec:privacy} for details). The resulting hierarchy is stored as a property graph (Figure~\ref{fig:property_graph}), in which every node carries a natural-language description and structured metadata. Each stage of the pipeline, as shown in Figure~\ref{fig:induce_module}, is additionally conditioned on user-provided context, described in~\S\ref{sec:user_context}.

\subsubsection{The Striving Hierarchy}
\label{sec:hierarchy}

\system{} draws on Activity Theory's hierarchical model of human behavior~\cite{kaptelinin2009acting, leontiev1978activity}, which distinguishes behavior by its purpose. Activity Theory's original formulation spans three levels: operations (automated responses to conditions), actions (goal-directed behaviors), and activities (motive-driven patterns). We extend this hierarchy upward with Emmons' concept of personal strivings~\cite{emmons1986personal}, which represent ongoing pursuits that organize multiple activities and persist beyond any single project. We find this extension necessary as Activity Theory's highest level (activities) remains tied to specific contexts and timescales, whereas strivings capture goals a person keeps returning to across contexts. 

We model \system{} around a means-ends hierarchy~\cite{kruglanski2018theory} where the same action can contribute to multiple activities, and the same activity to support multiple strivings. \system{} constructs this hierarchy through prompt chaining, drawing on the in-context learning capabilities of LLMs~\cite{brown2020language, xie2021explanation}. In this chain, each level's structured output becomes the input, and the conditioning context, for the level above. Our architecture is inspired by prior work on LLM-based concept induction and clustering~\cite{wang2023goal, zhong2023goal, zhong2022describing, lam2024concept}, which we extend to a chained hierarchical pipeline~\cite{jorke2025gptcoach, wu2022ai, shaikh2024rehearsal}.

\noindentparagraph{\textbf{Operations.}}
Operations are the lowest level of the hierarchy and describe atomic behavioral interactions such as a click, a scroll, or a keystroke. For instance, a screenshot of the first author browsing YouTube was transcribed as \texttt{browsing YouTube search results for classical music}. Unlike the other levels, operations simply record what happens on screen and make no claim about why.

\looseness=-1 \noindentparagraph{Operation Inference.} The observer captures screenshots and bundles a set of contiguous frames into a single observation. A vision-language model (VLM) then transcribes each observation into a structured markdown description of visible screen content, conditioned on user context. From each transcription, one or more operations are extracted, each carrying structured metadata (see Appendix~\ref{app:prompt-ops}).  

\noindentparagraph{\textbf{Actions.}}
Actions are goal-directed sequences composed from contiguous operations, where the system first makes the intent visible. For example, three operations: \texttt{messaging a friend about a football game on iMessage}, \texttt{searching for `Gareth Bale retirement' on Google}, and \texttt{reading a biography of Gareth Bale on Wikipedia}, span three applications but share a common intent and are grouped into a single action: \texttt{Following up on a football conversation by researching Gareth Bale's career}. We define the action boundaries between operations based on shifts in intent rather than changes in applications~\cite{mark2004const, mark2005no}.

\noindentparagraph{Action Inference.} \looseness=-1 Operations accumulate in a buffer until either a batch of operations has been collected or a period of inactivity elapses. The language model then segments the buffer into one or more actions. The segmentation prompt receives the buffered operations, recent actions, and local graph context as input and returns groups of operations belonging to a single goal-directed action. Each new action is then linked to its neighbors via three kinds of \emph{temporal relations}: a \texttt{follows} edge to the immediately preceding action, \texttt{co-occurs} edges to other actions extracted from the same buffer, and \texttt{overlaps} edges to recent actions whose time ranges intersect. The three are not mutually exclusive, so one pair may share all of them. These edges accumulate as observations arrive and become a key signal for grouping actions into activities and strivings.

\noindentparagraph{\textbf{Activities.}}
Activities represent recurring patterns of actions that share a common motive. They are broader than single tasks but narrower than a life domain. This level captures cross-application and cross-session synthesis. For example, over several days, the first author watched classical music performances on YouTube, ordered food on DoorDash, researched health-monitoring services, and followed election news on iMessage. Although these actions occurred at different times in different applications, the activity layer combined these into a single activity: \texttt{Managing homesickness and cultural identity through personal interests, food, news, music, and family messages}.

\noindentparagraph{Activity Inference.} When new actions are inferred, \system{} triggers a two-step propose-and-reconcile process. First, a proposal prompt takes as input new actions alongside the \texttt{follows}, \texttt{co-occurs}, and \texttt{overlaps} edges from the previous step, existing activities, and user context. It proposes possible activity groupings by considering semantic similarity, temporal co-occurrence, shared metadata, and graph-local structure. Then, a reconcile prompt integrates each candidate against the existing repository. It determines whether the candidate should match, revise, merge, or create a new activity. Activities themselves are further linked by their own temporal edges (for instance, when an activity consistently follows another, this insight could inform striving synthesis), and remain provisional until they recur across multiple batches.

\looseness=-1 \noindentparagraph{\textbf{Strivings.}}
Strivings represent ongoing personal pursuits in Emmons'~\cite{emmons1986personal} sense. While activities are organized around motives, each activity's motive is local to its cluster of actions. Consider two activities: \texttt{Managing homesickness and cultural identity} and \texttt{Researching family health monitoring services}. They involve different applications and immediate goals. At a broader level, however, they may reflect the same long-term pursuit. In this case, the striving layer synthesizes them into \texttt{Dedicated to preserving cultural identity and family connections while mitigating the isolation of life abroad}.

\noindentparagraph{Striving Inference.} Similar to activity inference, for each new activity, \system{} triggers a two-step synthesize and self-refine process to infer new strivings. First, a synthesis prompt takes the current set of reconciled activities alongside any existing strivings and user context. This step either maps activities to existing strivings or proposes new ones. A second prompt then critiques the synthesis against theory-grounded quality criteria, checking for redundancy and properties drawn from Emmons~\cite{emmons1986personal}. This self-refine pass~\cite{madaan2023self} can merge redundant strivings, split broad ones into multiple strivings, or recalibrate confidence scores. %The synthesis prompt is invoked once activities have begun stabilizing (after a warmup batch), with a twenty-four-hour ceiling enforced as a fallback.

\subsubsection{User-Provided Context}
\label{sec:user_context}

The same behavior can express fundamentally different pursuits depending on the user's circumstances and understanding. However, these circumstances and a user's self-understanding of a situation are hard to infer just from the screen. \system{} therefore elicits a lightweight self-description from the user before the observation begins. This self-description includes twelve  questions that cover current roles, daily routines, demands on time and energy, sources of stress, recent life changes, and priorities across six life domains: work, relationships, personal growth, health, finances, and education (full questions in  Appendix~\ref{app:onboarding}). Answers to these questions reveal information that cannot be reliably inferred from screenshots alone, such as caregiving responsibilities or ongoing transitions. More importantly, the elicitation does not ask users to specify their goals. The system still does the inferential work of constructing  strivings from observed behavior; however, the user's self-description changes \textit{how} the system interprets what it observes. We treat this onboarding context as a requirement for a system that aims to model what someone is working toward over the long term, and we evaluate its independent contribution to striving quality in \S\ref{sec:eval_induce}. User-provided context is also dynamic; as \system{} accumulates observations, users can revise their self-description when circumstances change.

\subsection{The Edit Module}
\label{sec:edit}

\begin{figure*}
    \centering
    \includegraphics[width=\textwidth]{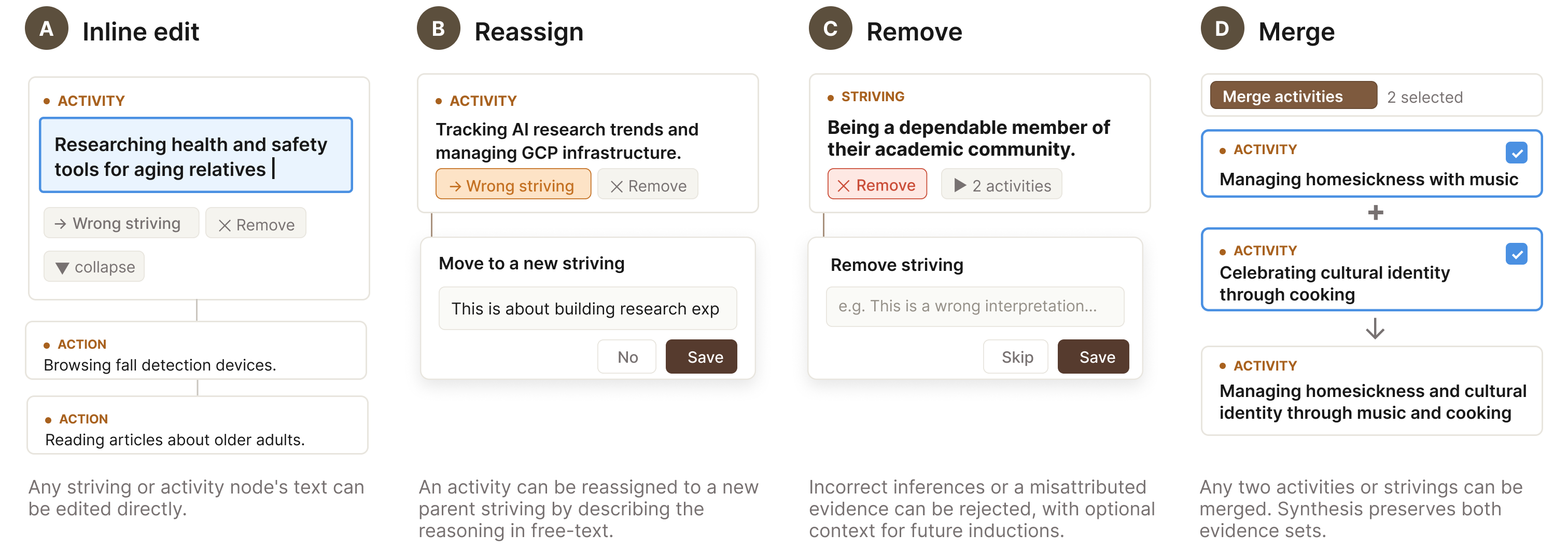}
    \caption{\system{} exposes four operations for editing the hierarchy:  (A) inline edit, (B) reassign,  (C) remove, and (D) merge. All  edits persist as constraints on subsequent induction cycles 
(\S\ref{sec:edit}).}
    \label{fig:edit_operations}
\end{figure*}

The Edit module allows the user to bring the hierarchy into alignment with their own self-understanding. This mechanism enables the system to resolve ambiguities that context alone cannot settle. Because the Induce module produces intermediate representations at four levels of abstraction, the Edit module surfaces strivings, activities, and the actions that ground them as editable targets, with the edit operations available at each level calibrated to the kinds of disagreements a user is likely to have there (Figure~\ref{fig:edit_operations}). Users can intervene wherever their disagreement lives, rather than reconstructing meaning from raw screenshots. The act of inspecting the hierarchy, tracing from a striving down through activities and actions to the screenshots that ground them, can itself prompt reflection on how daily behavior connects to broader pursuits. We return to these effects in our evaluation~(\S\ref{sec:eval_edit}).

\looseness=-1 For example, the first author noticed that \system{} had grouped Slack notifications about IRB logistics and paper coordination under an activity labeled \texttt{managing lab community engagement, technical awareness, and professional presence}, supporting a striving about \texttt{building social capital and professional visibility within the HCI community}. However, in this case, the first author was doing administrative work, not trying to build a community. Rather than locating the incorrect screenshots among hundreds, selecting the activity node showed the supporting evidence was IRB logistics and paper coordination. Removing this activity triggered \system{} to reassign the orphaned actions to a research activity and merge the two strivings into one: \texttt{establishing themselves as a reliable member of the HCI community through rigorous research, active mentorship, and peer support}.

\looseness=-1 \subsubsection{Editing Operations.} The Edit module exposes four operations through direct manipulation or natural-language input, each available at the levels where it applies (Figure~\ref{fig:edit_operations}). \textit{Inline edit} lets a user rewrite the text of a striving or activity when its synthesis is right but its description does not match how the user understands themselves. \textit{Reassign} moves an activity to a different parent striving. \textit{Remove} drops a striving, activity, or action whose evidence is wrong or unrepresentative. \textit{Merge} combines two strivings or two activities the system kept separate, preserving the union of their evidence. Operations are not surfaced as editable because they are descriptive records of screen events rather than inferences about purpose (\S\ref{sec:hierarchy}).

\looseness=-1 \subsubsection{Edit Feedback Loop.} When a user makes an edit, changes to the hierarchy are shown immediately in the interface, and relevant pipeline stages are re-triggered. Inline editing, merging, or reassigning activities across strivings re-runs striving synthesis; removing a striving orphans its child activities, which are reassigned by the activity inference step. Edits persist as constraints on all subsequent inductions. Edited labels are flagged \texttt{[user-edited]} in the property graph and preserved verbatim; reassigned activities carry \texttt{[user-reassigned]} flags. When the pipeline re-runs, these flags are surfaced to every downstream induction as constraints the model must respect, while the system enforces---at both the prompt level and through write-time guards in the pipeline that block any update that would relabel, merge, or delete a flagged entity---that user-edited nodes are never modified without explicit user action~\cite{hirsch2017designing, horvitz1999principles, lyons2021conceptualising}. Each induction cycle thus reflects both what the system observed and what the person knows about themselves (see Appendix~\ref{app:prompt-user-constraints} for the constraint block injected into each prompt).

\subsection{Privacy, Data Security, and Implementation}
\label{sec:privacy}

We designed \system{}'s data practices around the principle that the user should retain control over what is observed, what is inferred, and where their data is stored. This study was approved by our institution's IRB and underwent a data risk assessment by our institution's privacy office.

\subsubsection{User-Facing Controls.} \system{}'s dashboard lets users exclude specific applications and websites from capture, and users can pause recording at any time. During editing~(\S\ref{sec:edit}), users can remove individual screenshots from the evidence supporting any node.

\subsubsection{Automated Privacy Filters.} \label{sec:system_auto_privacy}
In order to strengthen \system{}'s user-facing controls,  we implement a two-stage automated filter. First, a local OCR model checks each screenshot against a library of personally identifiable information (PII) patterns (e.g., credit card numbers, API keys, health record identifiers, etc.) and deletes matching frames before they enter the pipeline. Second, based on prior work~\cite{shaikh2025creating}, a contextual integrity audit~\cite{nissenbaum2004privacy} uses the system's own inferences to model the user's disclosure norms, e.g., a banking login page would be blocked if the system has never observed the user sharing financial information.

\subsubsection{Implementation.} \system{} runs on macOS, with the inference pipeline and property graph storage written in Python and the Observer application and editing interface in Node.js.  All screenshots and inferences are stored locally on the user's device. Inference calls are routed through Google Cloud Platform under our institution's enterprise cloud agreement, which covers all data risk levels and prevents the use of our data for model training. All data was deleted from participants' devices at the end of the study. 
\section{Applications of the Striving Hierarchy}

\looseness=-1 Once constructed, the striving hierarchy functions as a general-purpose API for downstream applications. The hierarchy is stored as a property graph (\S\ref{sec:induce}) that applications can query at any level of abstraction, traverse to connect immediate behavior to long-term purpose, and subscribe to for updates when new strivings emerge or confidence scores shift. To illustrate the breadth of applications this API enables, we built two prototypes: a proactive agent that grounds suggestions in long-term purpose rather than immediate task context, and a behavioral coach that detects tensions between competing strivings and proposes interventions informed by the full graph.

\subsection{Proactive Assistance Across Strivings}

Proactive agents are systems that anticipate what a user needs and act on their behalf~\cite{maes1994agents}. Existing proactive agents typically operate at the task level, and as a result produce task-bound suggestions, e.g., recommending a movie when a user is browsing entertainment or suggesting a calendar slot during a scheduling conversation~\cite{lu2024proactive, shaikh2025creating}. The striving hierarchy enables a different kind of proactive support that is grounded not just in what the user is doing in the moment but also in the long-term goals their actions serve.

The first author was browsing a family health and safety monitoring service late at night as they were wrapping up their research work. A baseline system that generated suggestions from user context and recent behavior, without the striving hierarchy, proposed: ``Compare products or bookmark the AI features for future reference.'' While reasonable, this suggestion is myopic and anchored to what the user is doing in the moment. Instead, \system{} traversed the striving hierarchy and found that the browsing was related to the striving of \texttt{dedicated to preserving cultural roots and family connections to home country while mitigating the isolation of life abroad} that was in tension with an approaching deadline under their striving of \texttt{navigating the administrative and instructional requirements of their PhD to maintain their institutional standing and funding}. Because \system{} knew how the immediate actions related to the author's long-term goals and how those goals competed for time, it proposed the following suggestion to resolve the conflict: ``Reserve 30 minutes before your next assignment to complete the evaluation of the family safety monitoring service and send a brief update to your parents about it.'' \system{} reasons over a conflict between two long-term pursuits and grounds the intervention in an inferred motivation, i.e., the author's relationship to their parents.  

\subsection{Behavioral Coaching Grounded in the Hierarchy}

While behavior change interventions are most effective when they connect to goals the person actually holds~\cite{jorke2025bloom}, most deployed systems deliver generic, decontextualized intervention prompts~\cite{jorke2025gptcoach}. To address this gap, we built a coaching prototype that uses motivational interviewing (MI)~\cite{miller2012motivational} technique grounded in the hierarchy's evidence. At the start of each chat session, the coach identifies strivings with increasing or decreasing momentum, selects an MI strategy suitable for the user's trajectory, and grounds its opening message in specific behavioral evidence from the graph.
 
\looseness=-1 When the first author interacted with the coach before the UIST deadline, it opened by noting momentum toward \texttt{establishing yourself as a high-impact researcher} but declining time allocated toward \texttt{preserving meaningful connections}: ``The parts of your life that ground you are being squeezed into the margins.'' When the author replied that they were feeling stressed, the coach surfaced the author's pattern of late night work sessions that heightened feelings of isolation. Drawing on evidence that the author was already strategically streamlining logistical friction, the coach proposed repurposing that pattern: ``What if you carved out 15 minutes before your late-night food order to call someone back home? It might transform that window from high-stress debugging into a moment that addresses your feelings of isolation.'' The coach could produce such responses because the hierarchy let it detect tensions between strivings and serve one pursuit without disrupting another.
 
\section{Evaluation}
\label{sec:evaluation}
 
\system{}'s architecture contributes two modules: an \textit{Induce} module that progressively abstracts observations into a striving hierarchy, and an \textit{Edit} module that lets users trace and reshape the system's inferences. We evaluate each in turn. For the \textit{Induce} module, we ask whether the pipeline produces strivings that people recognize as accurate and representative of their lives, and whether hierarchy and user-provided context each contribute to this quality~(\S\ref{sec:eval_induce}). For the \textit{Edit} module, we ask whether the intermediate representations the Induce module constructs, the operations, actions, and activities linking observations to strivings, serve as a useful surface for co-creation ~(\S\ref{sec:eval_edit}). We deploy \system{} with $N=14$ participants over one week of naturalistic computer use, then assess the resulting strivings through blind ratings and an editing session in a lab visit.
 
\subsection{Participants}
\label{sec:participants}

We recruited 19 participants through word-of-mouth and internal messaging channels at our institution. Five did not complete the study or had insufficient data for the pipeline to stabilize activities and synthesize strivings. We required at least 500 processed screenshots, corresponding to roughly one hour of active screen time at our sampling rate. This yielded $N=14$ participants (see Appendix~\ref{app:participants} for participant background information) for all reported analyses (median 1,498 processed screenshots; range 765--9,435).

\subsection{Deployment}
\label{sec:deployment}
 
Participants installed \system{} on their personal MacBook and filled out a brief onboarding form to provide the necessary user context. The system then ran for seven days, capturing screenshots of user interactions. 
Participants consented to screenshots from their personal computer being processed by the Gemini API, and were fully informed of the system's privacy features and protections.
The full system and both ablations~(\S\ref{sec:induce_ablations}) ran in parallel on the same observation stream, but the participants were not shown the inferences as they were being generated. After a week of use, participants attended an hour-long lab session comprising striving ratings~(\S\ref{sec:eval_induce}) and editing tasks~(\S\ref{sec:eval_edit}), followed by a semi-structured interview.

\subsection{Evaluating the \textit{Induce} Module}
\label{sec:eval_induce}

The \textit{Induce} module has two key elements. The hierarchical structure organizes observations into a four-level representation, and user-provided context shapes how that hierarchy interprets what it observes. We compare the full system against two ablations to evaluate whether each contributes to striving quality.

\subsubsection{Ablations.}
\label{sec:induce_ablations}
 
\begin{itemize}[leftmargin=*]
    \item \textbf{\system{}}. The full hierarchical pipeline, seeded with the onboarding context.
    \item \textbf{\system{} $-$ \{User Context\}}. The full hierarchical pipeline, but without access to the onboarding context. 
    \item \textbf{\system{} $-$ \{Hierarchy, User Context\}}. The hierarchical pipeline is removed entirely. Rather than abstracting observations through operations, actions, and activities, this ablation synthesizes strivings from the observations directly using a sliding window of past observations that fit into a model's context window, without intermediate structure or user context.
\end{itemize}
 
Comparing \system{} to \system{} $-$ \{User Context\} holds hierarchy constant and varies context, isolating whether the person's onboarding input improves individual striving quality. Comparing \system{} $-$ \{User Context\} to \system{} $-$ \{Hierarchy, User Context\} removes the hierarchical pipeline entirely, testing whether the architecture produces a more representative set than direct inference from observations.

\subsubsection{Measures.}
\label{sec:induce_measures}

\looseness=-1 The goal of striving induction is to produce strivings that are individually well-formed and collectively representative of what the person is working toward.  We evaluate strivings at two levels: \textit{individual striving quality} and \textit{set-level representativeness}.

\textit{Individual striving quality} captures how well-formed a striving is. For each ablation, participants rated each system-generated striving on four dimensions: \textit{accuracy} (factually grounded in observed behavior), \textit{alignment} (reflects the person's priorities, not merely what they happened to do), \textit{abstraction} (described at the level of an ongoing pursuit, neither too specific nor too vague), and \textit{characteristicness} (captures a recurrent pattern in the person's life, not a one-time task), each on a $-3$ to $+3$ Likert scale. These dimensions operationalize Emmons'~\cite{emmons1986personal} criteria that strivings are recurring, meaningful, and characteristic of the individual.

\textit{Set-level representativeness} captures the quality of the set of strivings taken together. For each ablation, participants viewed the complete set of strivings and rated them collectively on three dimensions: \textit{coverage} (coverage of what the person is working toward), \textit{discovery} (discovery of goals not previously considered), and \textit{motive} (reflects what matters, not just what the person happened to do). Participants were also asked to think aloud while answering an open-ended question justifying their ratings. All surveys and questions are included in Appendix~\ref{app:eval-questions}.
 
\looseness=-1 Because each participant rated multiple strivings across multiple ablations and measures, we estimate effects with linear mixed-effects models. We include the model specifications and the reliability statistics that support pooling dimensions within each model in Appendix~\ref{app:stats}. 

\begin{figure}[t]
    \centering
    \includegraphics[width=\linewidth]{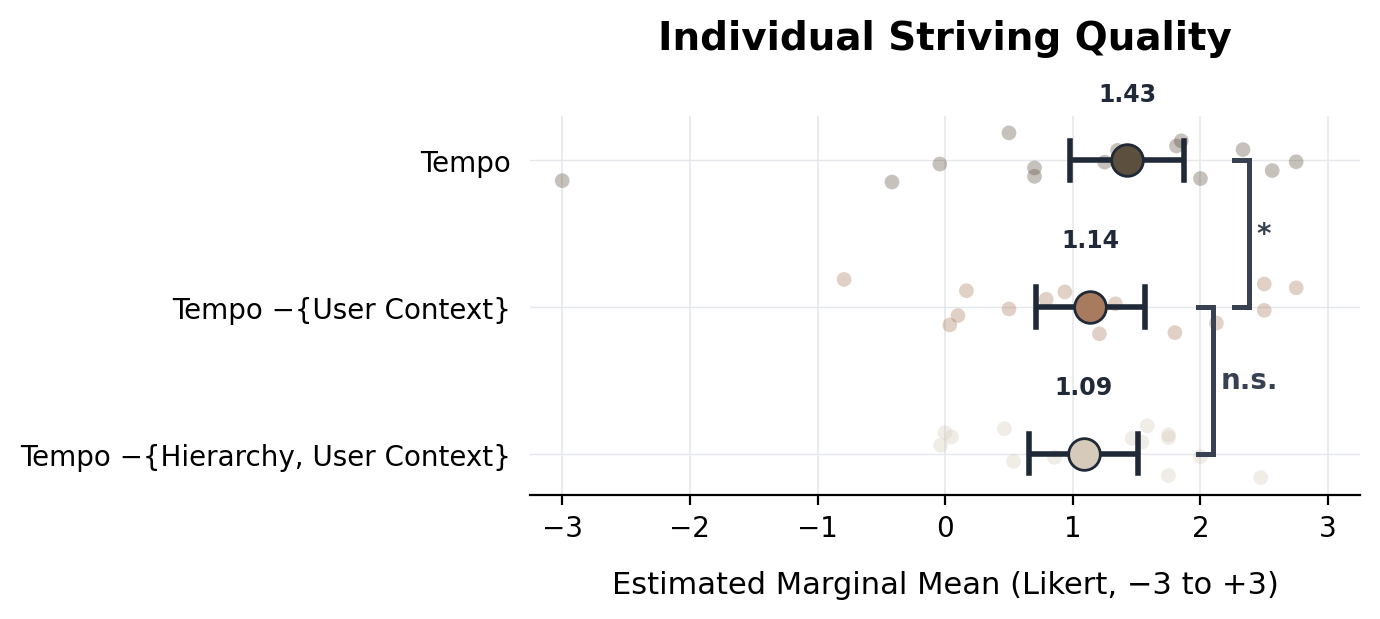}
    \\[4pt]
    \includegraphics[width=\linewidth]{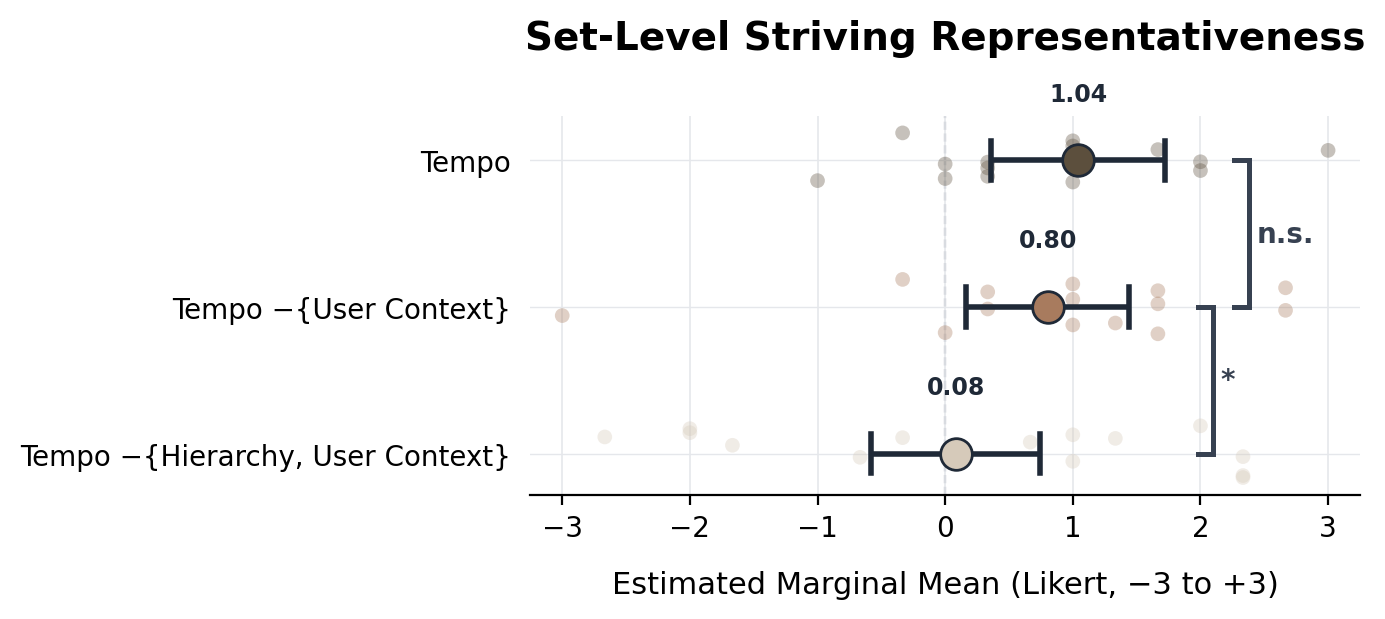}
    \caption{Estimated marginal means for striving quality. Top: individual striving quality pooled across four dimensions. User-provided context produced significantly higher per-striving ratings ($p<0.05$). Bottom: set-level representativeness pooled across three dimensions. Hierarchical structure produced significantly higher set-level ratings ($p<0.05$). Error bars show $\pm$ 1.96 SE of the marginal means; significance is assessed via mixed-effects model contrasts (see~\S\ref{sec:induce_measures}).}
    \label{fig:striving_quality}
\end{figure}

\subsubsection{Results.}
\label{sec:induce_results}
 
\noindentparagraph{\textbf{User-provided context improves striving quality.}}
 
User-provided context significantly improved individual striving quality: \system{} vs.\ \system{} $-$ \{User Context\} ($\beta = 0.290$, $p < 0.05 $). The effect of hierarchy alone was not significant at the individual striving level: \system{} $-$ \{User Context\} vs.\ \system{} $-$ \{Hierarchy, User Context\} ($\beta = 0.049$, $p = 0.679$). Figure~\ref{fig:striving_quality} shows estimated marginal means from the model.
 
A strong majority of strivings were rated positively ($\geq 1$) for accuracy across all ablations (\system{}: 80\%, \system{} $-$ \{User Context\}: 79\%, \system{} $-$ \{Hierarchy, User Context\}: 79\%), consistent with prior findings that non-hierarchical architectures produce individually accurate inferences~\cite{shaikh2025creating}. Rather, the contributions of context emerged in how well strivings reflected the person: \system{}'s advantage over \system{} $-$ \{User Context\} was most evident on \textit{alignment} ($M = 1.46$ vs.\ $M = 1.16$; 87\% vs.\ 73\% rated positively).

\noindentparagraph{\textbf{Hierarchy produces a more representative set of strivings.}}
 
Hierarchical structure significantly improved set-level representativeness: \system{} $-$ \{User Context\} vs.\ \system{} $-$ \{Hierarchy, User Context\} ($\beta = 0.721$, $p < 0.05$). Adding user context at the set level did not produce a significant improvement: \system{} vs.\ \system{} $-$ \{User Context\} ($\beta = 0.239$, $p = 0.503$). The hierarchical ablations also produced fewer strivings per participant (\system{}: $M = 4.9$, $SD = 1.4$; \system{} $-$ \{User Context\}: $M = 6.1$, $SD = 1.1$; \system{} $-$ \{Hierarchy, User Context\}: $M = 6.6$, $SD = 1.3$), yet scored higher on representativeness, indicating that the hierarchy's consolidation of observations into fewer, broader goals is not lossy.

The gap was largest on \textit{motive} (``reflects what matters to me, not just what I happened to do''). \system{} $-$ \{User Context\} ($M = 0.79$) and \system{} ($M = 0.57$) both substantially outperformed \system{} $-$ \{Hierarchy, User Context\} ($M = 0.07$). The motive gap is notable because it captures exactly the distinction between systems that describe \textit{what} a person is doing and systems that recover \textit{why} they are doing it. Our results suggest that the hierarchical structure is what enables that shift.

To illustrate these differences, we include a sample of anonymized strivings from the ablations shared by a participant after the study. 

\begin{quote}
\textbf{\system{}:} \texttt{P13 is asserting creative and technical agency through independent game development to establish a professional identity beyond his formal academic curriculum.} \\
\textbf{\system{} $-$ \{User Context\}:} \texttt{P13 is building complex AI systems that move beyond simple chat, aiming to create truly autonomous and interactive narrative story worlds.} \\ 
\textbf{\system{} $-$ \{Hierarchy, User Context\}:} \texttt{P13 is honing the tactile ``juice'' and mechanical impact of his game through iterative play-testing and real-time debug analysis.}
\end{quote}

While a week-long deployment cannot establish whether strivings persist over the long term, it can show whether they capture recurring aspects of participants' lives. Participants rated the inferred strivings as characteristic of their lives ($M = 1.35$). Many of these strivings described pursuits that clearly extended beyond the observation period. For P2, \system{} inferred \textit{``driving the transformation of [a startup] from a concept into a functional physical and operational reality.''}; for P12, \textit{``pursuing a sense of purpose through civic contribution, specifically by mentoring others and monitoring development strategies for their [home country]'s growth''}. Nor were the inferred strivings limited to work. Across participants, they covered family caretaking, startup survival, research, physical health, and creative exploration.

\noindentparagraph{\textbf{``These are my inner thoughts.''}}
 
When the onboarding context aligned with observed behavior, \system{} produced strivings that participants recognized as deeply personal. P5 described them as feeling \textit{``delivered from someone who knew me,''} and P12 said: \textit{``How the hell did it even know this\ldots It literally said this was my sense of purpose. And it is!''}. Without context, hierarchy alone surfaced goals operating beneath conscious articulation. P8 noted these were \textit{``so much in the background---I don't explicitly think or talk about them,''} while P1 found that \system{} without context \textit{``uses descriptors that I would never use to describe myself if asked---but that I should consider using in the future.''}
 
Participants' descriptions of the difference between ablations echoed this pattern. Strivings from \system{} $-$ \{Hierarchy, User Context\} were perceived as anchored to specific actions rather than the broader purposes those actions serve: \textit{``more of these felt like things that I just do and don't align with my goals, like using a terminal or watching YouTube videos''} (P14). By contrast, participants valued the hierarchical ablation as validator of invisible effort. P1 reported: \textit{``A lot of times when I reflect on my work at the end of the week, I feel like I haven't done a lot of work. And the system provided that reassurance that I actually did a lot.''}
 
The onboarding context, however, was a double-edged sword. \system{} achieved the highest ceiling but introduced confident misattributions when context was over-weighted. P1 noted \textit{``I am not pivoting to a career as a software engineer,''} and P2 noted some strivings seemed \textit{``the interpretations of some mundane actions overweighted on my onboarding responses.''}

\noindentparagraph{\textbf{Errors and boundaries.}} Despite these strengths, participants identified recurring errors with \system{}. The system sometimes merged unrelated activities under a single striving because they shared surface features or co-occurred temporally (P13, P9). Requiring a minimum evidence threshold before surfacing a striving could reduce such false consolidations. Strivings also over-represented work activity, likely because the screen observer captured more professional usage than informal personal use. The full system sometimes over-indexed on the user context to generate confident strivings unsupported by behavioral evidence, suggesting that context needs to be more systematically weighted during striving generation.

\subsection{Evaluating the \textit{Edit} Module}
\label{sec:eval_edit}
 
The \textit{Induce} module constructs intermediate representations that link observations to strivings. The \textit{Edit} module makes this reasoning trace a surface for co-creation. Can users follow how the system arrived at its inferences and meaningfully reshape them? Contestability research has identified several properties that systems need to support meaningful challenge of algorithmic inferences, including legibility of the reasoning process~\cite{hirsch2017designing}, access to underlying evidence~\cite{kay2013creating}, and mechanisms for users to shape the representations a system holds about them~\cite{vaccaro2019contestability, lyons2021conceptualising}. We evaluate whether \system{}'s hierarchical trace supports these properties, and whether exposing it, as opposed to the raw observations it was built from, changes the editing experience.

\subsubsection{Procedure.}
 
Participants edited strivings using two interfaces (see Appendix~\ref{app:system-fig}) in counterbalanced order during the lab session. In the \textit{screenshot view}, participants edited all strivings generated by \system{} $-$ \{Hierarchy, User Context\}, each displayed alongside the raw screenshots used to generate it. Participants could edit striving text inline, remove or redistribute screenshots across strivings, remove strivings entirely, or merge multiple strivings into one. In the \textit{hierarchical view}, participants edited strivings from the full \system{}, each linked to its constituent activities as collapsible cards expanding to show actions and corresponding screenshots. Participants could perform the same editing operations as the screenshot view, but could also navigate and edit at any level of the hierarchy, including removing an action from an activity, removing or relabeling an activity, restructuring which activities fed into a striving, or removing a striving. In both interfaces, we asked participants to think aloud as they made edits, while logging all editing interactions: timestamped click events (categorized by target: screenshot, button, text input, navigation), edit operations, and total session duration.
 
Because the hierarchical view shows strivings from \system{} and the screenshot view shows strivings from \system{} $-$ \{Hierarchy, User Context\}, striving quality and evidence structure co-vary across views. The measures in this evaluation therefore focus on the editing \textit{process} rather than output quality, which was assessed in~\S\ref{sec:eval_induce}. 

\subsubsection{Measures.}
 
\begin{figure}
    \centering
    \includegraphics[width=\linewidth]{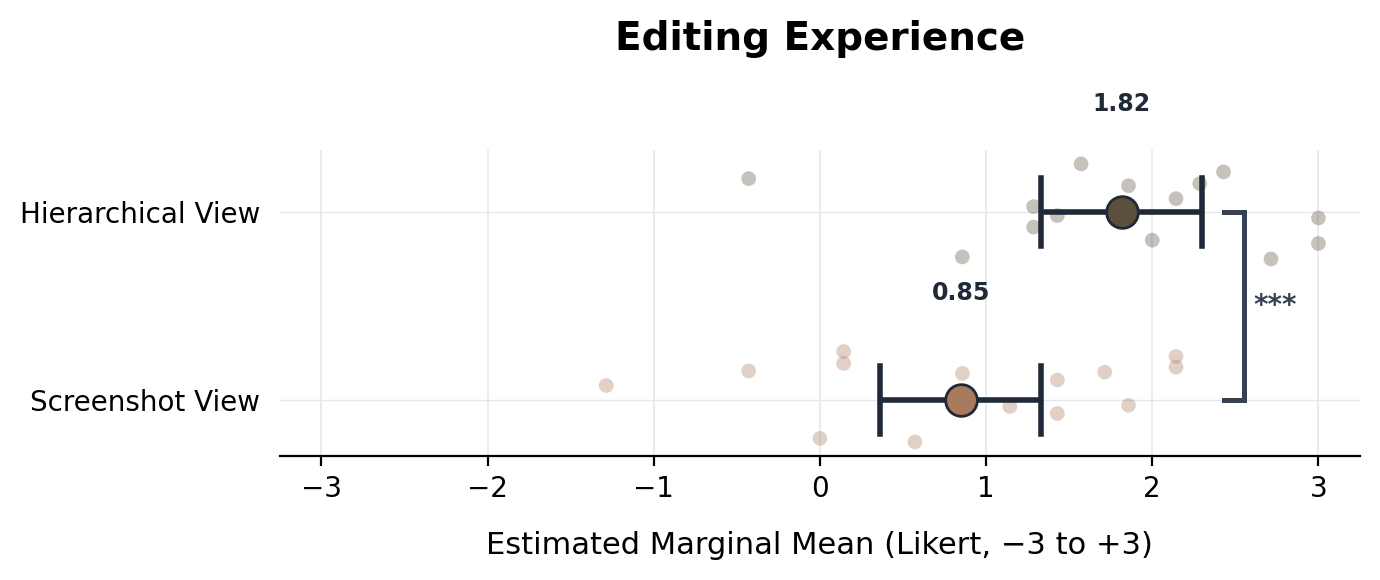}
    \caption{Estimated marginal means for the editing experience pooled across seven dimensions. The hierarchical view was rated significantly higher across all dimensions ($p<0.001$). Error bars show $\pm 1.96$ SE (\S\ref{sec:eval_edit}).}
    \label{fig:striving_edit}
\end{figure}

\looseness=-1 After editing in each interface, participants rated seven dimensions on $-3$ to $+3$ Likert scales, drawn from contestability and scrutable user models literature: \textit{transparency}~\cite{hirsch2017designing}, \textit{evidence utility}~\cite{kay2013creating, balog2019transparent}, \textit{control}~\cite{vaccaro2019contestability, lyons2021conceptualising}, \textit{agency}~\cite{horvitz1999principles}, \textit{confidence}, \textit{reflection}~\cite{fleck2010reflecting, bentvelzen2022revisiting}, and \textit{ease} (see Appendix~\ref{app:post-edit} for full set of questions). We analyzed ratings with the same mixed-effects approach (Appendix~\ref{app:stats}). Figure~\ref{fig:striving_edit} displays estimated marginal means from this model. 

\subsubsection{Results.}
 
\noindentparagraph{\textbf{The hierarchical trace transforms the editing experience.}}
 
The hierarchical view was rated substantially higher than the screenshot view ($\beta = 0.969$, $p < .001$). The strongest effects were on \textit{transparency} ($\Delta = +2.14$) and \textit{evidence utility} ($\Delta = +1.71$), the dimensions most directly tied to whether users could follow how the system arrived at its inferences. Participants rated the hierarchical view as providing sufficient transparency at 93\% compared to 43\% for screenshots, and 100\% rated the hierarchical evidence as useful compared to 71\% for screenshots.

\noindentparagraph{\textbf{The reasoning trace changes how participants navigate evidence.}}
 
Participants reported that the hierarchical structure let them follow how the system moved from observed behavior to higher-order goals. P6: \textit{``The hierarchy gives you more control-- otherwise you need to synthesize at the action level yourself.''} P10 found \textit{``text-based interpretations of the actions was a better medium for capturing what was actually happening on the screen.''} 

By contrast, participants found raw screenshots difficult to interpret as evidence for higher-order strivings. In the screenshot view, screenshot browsing accounted for 36.6\% of all clicks on average, with 6 of 14 participants spending more than 30\% of their interaction scrolling through evidence they could not connect to the strivings it supported. P10: \textit{``It was tough with the screenshots to understand the reasoning behind using certain screenshots as evidence for a particular goal.''} Without the reasoning trace, participants could identify \textit{that} a striving was wrong but not \textit{why}: \textit{``I don't know how to edit the goal based on what you've captured''} (P1).

\noindentparagraph{\textbf{Reshaping the reasoning trace.}}
 
Rather than simply accepting or rejecting, participants used the hierarchy to locate where the system's interpretation diverged from their own understanding. P6 observed: \textit{``Certain clusters can appear under different goals. But you can see here that there's one throughline throughout your life. This helped me understand that even at low-level things, why am I looking into these things is the product of the same life philosophy.''} P12 described the experience as \textit{``like a therapist looking at my behavior and analyzing what I'm doing and giving me a totally new insight.''}

\noindentparagraph{\textbf{Participants wanted the system to propose alternatives.}}
 
Despite high control and agency ratings, participants identified a gap when a striving was wrong. They often lacked the self-knowledge to specify a replacement. P2: \textit{``When I say there's a problem with this or it's wrong, I don't really know what the goal is myself, so maybe it could regenerate a couple options and I could choose from them.''} Even with the hierarchy showing the reasoning trace, participants wanted the system to go further, proposing candidate replacements rather than leaving them to author corrections from scratch.

\noindentparagraph{\textbf{Errors and boundaries.}}
 
Despite the transparency afforded by the hierarchy, 2 of 14 participants preferred the screenshot view, citing its simplicity and coverage: P5 noted it \textit{``required fewer button clicks and was just simpler to understand,''} and P8 found that showing more strivings \textit{``felt like it had higher coverage and therefore felt more accurate.''} Even among those who preferred the hierarchical view, participants noted it took longer (P5) and surfaced more structure than they wanted to engage with. P14 liked the collapsible cards but felt \textit{``the number of actions could be lower.''} These preferences suggest a design tension between reasoning transparency and cognitive cost. Exposing the full hierarchy aids diagnosis but risks overwhelming users who want to edit quickly.
 
Beyond this transparency tradeoff, P10 also reported a boundary on what is possible through behavioral inference alone: \textit{``my life goals are a lot broader\ldots You have to know stuff about my history and my past before I started using this laptop.''}

\section{Discussion}
\label{sec:discussion}
Our work points to a shift in user modeling from capturing what a person is doing in the moment to interpreting what they are working toward over the long term. Here, we reflect on what this shift implies for co-creative systems, along with limitations and directions for future work.

\noindentparagraph{\textbf{Co-creation as reflection.}}

\looseness=-1 Even before being prompted to edit a striving, when participants read \system{}'s inferences, they often reflected on what they had been doing and why. This led participants to trace their own behavior back to the higher-level goal and draw connections across their day they had not previously seen. When the evidence matched participants' own understanding, the reflection surfaced behavior that they had forgotten, while mismatched evidence prompted them to reconsider why they had been doing something they could not immediately connect to a long-term goal. Similar to prior work~\cite{jorke2023pearl, shaikh2025creating}, this tension between system interpretation and self-understanding was a rich source of reflection.

These responses suggest co-creative mechanisms might require interactions beyond direct editing. A conversational interface, for instance, could let users interrogate why the system grouped certain behaviors together or articulate goals the system has no behavioral evidence for. Participants who struggled to author replacement strivings from scratch would also benefit from a system that proposes candidate alternatives rather than leaving the burden of articulation entirely to the person. This interpretation is consistent with concurrent findings that people want ways to correct inferences that misrepresent them~\cite{monteiro2026llm}. 

\looseness=-1 Reviewing every inference becomes impractical as the hierarchy grows, so a mixed-initiative design could surface the hierarchy only when strivings conflict or the system is uncertain. In those moments, the aim would be reflection as well as correction. As noted in \S\ref{sec:rw_reflection}, this co-creation process also reverses the usual framing of the intention-behavior gap~\cite{faries2019int}. The mismatches participants found during editing make that gap visible without relying only on self-report.

\noindentparagraph{\textbf{Strivings as interpretive artifacts, not ground truth.}}

Our deployment showed that co-creation did not always ensure critical engagement from our participants. Several participants accepted the strivings without checking the behavioral evidence because the system had said something they found validating. One side of this dynamic is documented in sycophancy research, where models trained on human feedback drift toward what users want to hear~\cite{cheng2025social}. On the other side is the Barnum effect~\cite{snyder1977acceptance} where people read vague personality descriptions as uniquely their own. 

\looseness=-1 Despite this risk, we observed that anchoring strivings in a visible behavioral trace, rather than presenting them as standalone assertions, was important to make strivings scrutable. The hierarchical probe allowed participants to ask whether the behavior underneath warranted the interpretation. Still, the trace only helps if people return to it. Future systems could thus prompt periodic re-evaluation and ask users to revisit stale strivings or flag when new behavior diverges from established constraints. Inferred strivings should also be treated as interpretive artifacts that support self-understanding~\cite{pang2025interactive} and not as objective records of a person's goals.

\noindentparagraph{\textbf{Privacy and security risks.}}
 
Because \system{} infers strivings from ongoing observations of behavior, an attacker could try to manipulate what the system concludes a person is working toward. Prior work on user modeling has, in fact, demonstrated that spam email could maliciously rewrite user model propositions~\cite{shaikh2025creating}. The risk may be especially pronounced on shared devices, where unintended observations could influence the inferred hierarchy. The Audit module provides one layer of safeguard against such manipulation, but it does not by itself establish that the observation pipeline is robust to deliberate manipulation.

The inferred hierarchy could also introduce a separate security concern. Although an individual screenshot captures one moment, the hierarchy brings together weeks of behavior and organizes it into an account of what a person may be trying to achieve across different parts of their life. Access to it could therefore reveal not only what the person did, but also how the system interpreted their motives. The hierarchy may be more sensitive than any single observation from which it was constructed. Thus, future deployments should treat the hierarchy with the same protections as a credential. 

\noindentparagraph{\textbf{The observer effect: beyond technical privacy.}}

We designed \textsc{Tem\-po} to observe behavior passively, but several participants reported that the observation itself changed the behavior it aimed to capture. P2 mentioned working harder than usual, P4 actively suppressed distractions, and P11 blocked WhatsApp entirely. In each case, participants performed a version of themselves they wanted the system to see. This reactivity is well-documented in clinical and personal informatics contexts~\cite{kersten2017personal}, and when the system infers strivings from performed behavior, the resulting model may reflect an aspirational self rather than an actual one~\cite{higgins1987self}. 

However, this effect may also be transient. A few participants reported that awareness of the observer faded over the week, and one forgot the system was running entirely, asking about our privacy mechanisms only after the fact. But forgetting the system is running introduces its own problem. \system{} includes mechanisms to give users control over what is observed: pausing recording, an application exclusion list, local OCR-based PII filtering, and removing screenshots during editing. For a system designed to run continuously over weeks, privacy mechanisms that depend on active user engagement are brittle. Our week-long deployment offers initial evidence for the benefits of striving-level inference, while leaving open how those benefits should be weighed against the risks of continuous screen capture. Future systems could reduce these concerns by drawing on application APIs or other user-permissioned signals, though the right balance remains an open question as systems move from week-long studies to sustained deployment.

\noindentparagraph{\textbf{Limitations.}}

\looseness=-1 Our participant pool reflects a self-selection effect, i.e., they were already comfortable with AI tools, agreed to have their screens recorded for a week, and spent much of each day on a computer. These characteristics may have made participants more receptive to the inferred strivings than a broader population. All participants also used one machine for both work and personal life (Appendix~\ref{app:participants}), so our findings may not generalize when behavior is distributed across devices, further motivating future work on multi-source inference. We also did not quantify what the privacy audits removed or how participants responded to filtered versus retained evidence, which remain important questions for future work.

The inferred strivings also occasionally exhibited verbose and confident language that is characteristic of LLMs. Better prompting and grounding can reduce this problem, but interfaces should also help users tell well-supported interpretations from speculative ones~\cite{horvitz1999principles}. The system is also limited by what it can observe. Participants described relationships, health, and other offline parts of their lives as central to their goals, yet little of this context appeared in the captured data, so the hierarchy skewed toward structured, computer-based work. Future systems could allow users to contribute behavior from additional sources, e.g., wearable data, mobile activity, or self-reported journal entries, to broaden the observation window. Rather than collecting additional data in the background, the system could let people decide which parts of their lives should inform its interpretations. 

Our system also has some architectural limitations. Each induction cycle currently takes 5--10 minutes depending on the volume of buffered observations. Since strivings emerge over several days rather than from a single moment, this delay may be tolerable, but faster inference would let users review and correct interpretations while the relevant behavior is still recent. The pipeline also depends on several configurable values, including buffer size, graph-retrieval limits, stabilization thresholds, and confidence cutoffs, whose default values have not been robustly tested across different patterns of computer use. 

Finally, our evaluation spans one week of observation and a single editing session. Although participants recognized \system{}'s strivings as ongoing patterns in their lives, one week cannot establish that the inferred strivings are stable, nor can it reveal how they evolve with a user’s changing circumstances. We thus need longitudinal deployments to evaluate the temporal persistence of the strivings and whether edit constraints produce increasingly aligned strivings across multiple induction cycles.

\section{Conclusion}

This paper introduces striving co-creation, a process where a person and a system jointly construct a representation of the person's long-term life goals from everyday computer use. \system{} builds this representation as a hierarchy grounded in Activity Theory and Emmons' personal strivings framework and incorporates the person's corrections as constraints on future inferences. In a week-long deployment ($N=14$), participants found the strivings inferred by \system{} to be accurate and representative of their lives. Our ablation results show that user-provided context improves the quality of individual strivings, while hierarchical structure improves the representativeness of the overall set and supports editing by letting participants trace how the system arrived at an inference and reshape it. We find that a week of observation surfaces pursuits that extend well beyond the observation window, but assessing the stability of the strivings will require longer deployments. We hope that \system{} points toward systems that support people in their long-term pursuits while treating these pursuits as interpretations for negotiation rather than facts to be extracted.

% \section{Modifications}

% Modifying the template --- including but not limited to: adjusting
% margins, typeface sizes, line spacing, paragraph and list definitions,
% and the use of the \verb|\vspace| command to manually adjust the
% vertical spacing between elements of your work --- is not allowed.

% {\bfseries Your document will be returned to you for revision if
%   modifications are discovered.}

\begin{acks}
We thank the members of the Stanford HCI  group for their support on this project. We thank Danilo Symonette, Nava Haghighi, Jane E, Mikaela Angelina Uy, Michelle Lam, Jordan Troutman, Julia Markel, Dora Zhao, Parker Ruth, Elizabeth Childs, Michael Ryan, Michael Li, and Ralf Herbrich for their helpful discussions and thoughtful insights. We also thank Tonya Nguyen, Helena Vasconcelos, and Jennifer Wang for their feedback on the system, Vincent Eichhorn for contributions to the early prototype of the system, and Kit Ling Leong for help with the video figure. Our work is supported by the Stanford Institute for Human-Centered Artificial Intelligence (HAI) and the Hasso Plattner Foundation. 
\end{acks}

\bibliographystyle{ACM-Reference-Format}
\bibliography{base}

\appendix
\newpage
\section{Onboarding Survey Questions}
\label{app:onboarding}

Participants completed an onboarding survey at the start of the study. Responses were used to construct the user context provided to the pipeline (see Section~\ref{sec:user_context}).

\begin{enumerate}
  \item What are the main roles you play in your life right now?
  \item Walk me through what a typical weekday looks like for you.
  \item What takes up most of your time and energy right now---whether by choice or obligation?
  \item What are the main sources of stress or pressure in your life right now?
  \item Has anything significant changed in your life recently, or is anything about to change?
  \item In terms of your work or career, what matters most to you or concerns you most?
  \item In terms of your relationships, what matters most to you or concerns you most?
  \item In terms of your personal growth, what matters most to you or concerns you most?
  \item In terms of your health, what matters most to you or concerns you most?
  \item In terms of your finances, what matters most to you or concerns you most?
  \item In terms of your education, what matters most to you or concerns you most?
  \item Is there anything else about your life or situation that would help us understand your computer use?
\end{enumerate}

\section{Evaluation Questions}
\label{app:eval-questions}

\subsection{Individual Striving Evaluation}
\label{app:goal-eval}

Each inferred striving was rated on a 7-point Likert scale ($-3$ = Strongly Disagree, $+3$ = Strongly Agree):

\begin{enumerate}
  \item \textbf{Accuracy.} ``This statement accurately describes something I actually do or try to do.''
  \item \textbf{Alignment.} ``This statement captures something that currently matters to me.''
  \item \textbf{Abstraction.} ``This statement captures what I'm trying to achieve, not a specific way to get there.''
  \item \textbf{Characteristicness.} ``This is an ongoing pattern in my life, not a one-time task.''
\end{enumerate}

\subsection{Set-Level Evaluation}
\label{app:set-eval}

After rating individual strivings, participants evaluated each condition's full set on a 7-point Likert scale ($-3$ = Strongly Disagree, $+3$ = Strongly Agree):

\begin{enumerate}
  \item \textbf{Coverage.} ``This set of statements captures what I am trying to achieve.''
  \item \textbf{Discovery.} ``This set surfaced goals or connections I hadn't previously thought about.''
  \item \textbf{Motive.} ``This set reflects what matters to me, not just what I happen to do.''
\end{enumerate}

\noindent Participants were also asked to provide an open-ended explanation of their rating.

\subsection{Post-Edit Survey}
\label{app:post-edit}

After editing each condition's strivings, participants rated the editing experience on a 7-point Likert scale ($-3$ = Strongly Disagree, $+3$ = Strongly Agree):

\begin{enumerate}
  \item \textbf{Control.} ``I felt in control of shaping this set of goals to match my actual life.''
  \item \textbf{Agency.} ``I could meaningfully change things that were wrong, not just accept or reject them.''
  \item \textbf{Transparency.} ``I could understand why the system generated each goal.''
  \item \textbf{Evidence Utility.} ``The evidence the system showed me helped me decide whether to keep or change a goal.''
  \item \textbf{Reflection.} ``Looking at these goals made me think about my priorities in a new way.''
  \item \textbf{Confidence.} ``After editing, I am confident this set accurately represents my goals.''
  \item \textbf{Ease.} ``The editing process was straightforward.''
\end{enumerate}

\subsection{Statistical Analysis}
\label{app:stats}

We use linear mixed-effects models to estimate overall effects while accounting for the repeated-measures structure of the ratings. For the \textit{Induce} evaluation (\S\ref{sec:eval_induce}), we fit separate models for per-striving quality and set-level representativeness, each with ablation, number of inferred strivings, and measure as fixed effects and random intercepts for participants. The four per-striving dimensions are internally consistent (Cronbach's $\alpha = 0.86$), as are the three set-level dimensions ($\alpha = 0.79$), supporting pooling them in their respective models. We additionally report per-measure means in the main text to show where effects concentrate. 

\looseness=-1 For the \textit{Edit} evaluation (\S\ref{sec:eval_edit}), we fit a mixed-effects model pooling across all seven dimensions (Cronbach's $\alpha = 0.81$), with random intercepts for participants, and interface type and measure as fixed effects. 

Bootstrapped 95\% confidence intervals for the reported contrasts are: \system{} vs.\ \system{} $-$ \{User Context\} on individual striving quality ($\beta = 0.290$, CI $[0.026, 0.571]$), \system{} $-$ \{User Context\} vs.\ \system{} $-$ \{Hierarchy, User Context\} on individual striving quality ($\beta = 0.049$, CI $[-0.198, 0.263]$), \system{} $-$ \{User Context\} vs.\ \system{} $-$ \{Hierarchy, User Context\} on set-level representativeness ($\beta = 0.721$, CI $[0.099, 1.343]$), \system{} vs.\ \system{} $-$ \{User Context\} on set-level representativeness ($\beta = 0.239$, CI $[-0.454, 0.930]$), and the hierarchical view vs.\ the screenshot view on the editing experience ($\beta = 0.969$, CI $[0.613, 1.321]$).

\section{Semi-Structured Interview Questions}
\label{app:interview-protocol}

After completing the editing tasks (\S\ref{sec:eval_edit}), we conducted a semi-structured interview using the following question set. We also asked follow-ups as needed to clarify or probe further.

\noindentparagraph{\textbf{Inference quality}}
\begin{enumerate}
  \item In either editing view, were there goals where you felt like the system really captured something about you? What about ones that felt off?
  \item Did either version connect things from different parts of your life, like work and personal, or health and hobbies? How did that land for you?
  \item Were there any goals that surprised you? Things you're actually pursuing but hadn't thought of as a goal?
  \item Did the system surface anything you know is true about you but wouldn't have thought to write down yourself?
\end{enumerate}

\noindentparagraph{\textbf{Editability}}
\begin{enumerate}\setcounter{enumi}{4}
  \item When you disagreed with a goal the system generated, could you figure out where it went wrong? Was that different between the two views?
  \item Can you walk me through one goal you changed or rejected? What was wrong and how did you fix it?
  \item Did looking at the screenshots and activities underneath a goal ever change your mind about it. For example, either keeping something you would have rejected, or rejecting something that initially looked right?
\end{enumerate}

\noindentparagraph{\textbf{Comparative reflection and self-discovery}}
\begin{enumerate}\setcounter{enumi}{7}
  \item You saw your goals organized two different ways. Was one easier to work with? Why?
  \item If you had sat down a week ago and written out your life goals from scratch, do you think you would have written the same things the system came up with? What would have been different?
\end{enumerate}

\section{Abridged Prompts}
\label{app:prompts}

We present abridged versions of the key pipeline prompts. Placeholders are shown in \texttt{\{braces\}}.

\subsection{User Constraint Block}
\label{app:prompt-user-constraints}

The \texttt{\{user\_constraints\}} placeholder is injected into the activity proposal (\S\ref{app:prompt-activity-propose}), activity reconciliation (\S\ref{app:prompt-activity-reconcile}), striving synthesis (\S\ref{app:prompt-striving-synthesize}), and self-refine (\S\ref{app:prompt-striving-refine}) prompts. It is rendered dynamically from the property graph at each induction cycle. For every goal and activity carrying user metadata, the block emits one line per flag:

\begin{quote}
\texttt{\# User constraints}\\
\texttt{\{user\_name\} has edited, locked, or annotated the following entities.}\\
\texttt{Respect these constraints:}\\
\texttt{- [locked] goals: Do NOT merge, delete, or substantially alter.}\\
\texttt{- [user-edited] goals: Keep the exact label verbatim.}\\
\texttt{- [user-reassigned] activities: Keep with their assigned goal.}\\
\texttt{- Annotations provide privileged context --- weight above behavioral inference.}

\texttt{- Goal ID:\{id\} | \{label\} | [locked] --- do not merge, delete, or substantially alter}\\
\texttt{- Goal ID:\{id\} | \{label\} | [user-edited] --- preserve exact label verbatim}\\
\texttt{- Goal ID:\{id\} | \{label\} | [endorsed] --- user confirmed this goal is accurate, keep it}\\
\texttt{- Goal ID:\{id\} | \{label\} | [rejected] --- user says this goal is wrong, do not recreate it}\\
\texttt{- Activity ID:\{id\} | \{label\} | [user-reassigned] --- respect current goal assignment}\\
\texttt{- Activity ID:\{id\} | \{label\} | annotation (\{type\}): "\{text\}"}
\end{quote}

An additional \texttt{\{user\_review\_edits\}} block, used by striving synthesis and self-refine, summarizes the most recent review session: which goals or activities the user accepted, removed, edited, or merged, and whether screenshots were removed from any node.

\noindent\textit{Note: The constraints above are enforced not only at the prompt level but also as write-time guards in the reconciliation pipelines that block any update violating a flag. Even if the language model returns a decision that would relabel, merge, or delete a \texttt{[user-edited]} or \texttt{[locked]} entity, the corresponding writes are guarded. In such a case, a relabel is dropped, a merge into a locked target is downgraded to a match, and a locked source is skipped from any merge group. The prompt instructs the model to behave and the code enforces these requirements. Together, these blocks and the accompanying guards are how the Edit module's interventions (\S\ref{sec:edit}) propagate as constraints into every subsequent induction cycle.}

\subsection{Screen Observation}
\label{app:prompt-screen}

The screen observer captures bundles of contiguous screenshots and feeds them to a vision-language model in two modes. The transcription mode produces the verbatim markdown that becomes \texttt{\{observation\_text\}} in operation extraction (\S\ref{app:prompt-ops}); the summary mode produces a complementary action-focused description of the same frames.

\noindentparagraph{\textbf{Transcription mode.}}

\begin{quote}
\texttt{Transcribe in markdown ALL the content from the screenshots of the user's screen.}

\texttt{NEVER SUMMARIZE ANYTHING. You must transcribe everything EXACTLY, word for word, but don't repeat yourself.}

\texttt{ALWAYS include all the application names, file paths, and website URLs in your transcript.}

\texttt{\{user\_context\}}

\texttt{Create a FINAL structured markdown transcription.}

\texttt{Screenshots: \{screenshots\}}
\end{quote}

\noindentparagraph{\textbf{Summary mode.}}

\begin{quote}
\texttt{Provide a detailed description of the actions occurring across the provided screenshots. The screenshots are in the order they were taken.}

\texttt{Include as much relevant detail as possible, but remain concise.}

\texttt{Generate a handful of bullet points and reference \textit{specific} actions the user is taking.}

\texttt{Keep in mind that the content on the screen is what the user is viewing. It may not be what the user is actively doing or what they believe, so practice caution when making assumptions.}

\texttt{Screenshots: \{screenshots\}}
\end{quote}

\subsection{Operation Extraction}
\label{app:prompt-ops}

\begin{quote}
\texttt{You are given a transcription of what is visible on a user's screen. Extract operations---an important atomic action the user is performing.}

\texttt{For each operation, provide:}\\
\texttt{- text: natural-language description}\\
\texttt{- confidence (1--10): how certain this operation occurred}\\
\texttt{- decay (1--10): how long this remains meaningful}\\
\texttt{- reasoning: why you extracted this operation}

\texttt{Structured metadata:}\\
\texttt{- tool\_kind: \{editor, messaging, browser, docs, calendar, data\_analysis, other\}}\\
\texttt{- social\_target: who the operation is directed toward \{advisor, lab, collaborators, friends, family, null\}}\\
\texttt{- rule\_tags: explicit rules/norms (e.g., deadline, compliance, admin requirement)}\\
\texttt{- automaticity\_hint: \{likely\_habitual, \\likely\_deliberate, unclear\}}\\
\texttt{- affect\_hint: \{stressed, rushed, relaxed, neutral, null\}}

\texttt{User context: \{user\_context\}}\\
\texttt{Screen transcription: \{observation\_text\}}
\end{quote}

\subsection{Action Segmentation}
\label{app:prompt-actions}

\begin{quote}
\texttt{You are given a sequence of operations [1..N]. Segment them into goal-directed actions. Segments must be contiguous, \\non-overlapping, and cover all operations.}

\texttt{An ACTION is a coherent task the user would describe in one sentence. Use \{user\_feedback\} to avoid supporting rejected goals/activities and to align with endorsed or edited ones. Don't over-segment for minor tweaks within the same task. Reference specific applications, documents, projects, or names.}

\texttt{Per-action metadata:}\\
\texttt{- object\_label: main object/problem}\\
\texttt{- outcome\_type: \{produce\_artifact, communicate, plan\_organize, learn\_explore, monitor\_check, other\}}\\
\texttt{- domain: \{research, teaching, admin, personal\_\\life, health, other\}}\\
\texttt{- community: \{lab, teaching\_team, students, family\_friends, null\}}\\
\texttt{- engagement\_state: \{sustained\_focus, fragmented, idle, rapid\_switching\}}\\
\texttt{- cognitive\_mode: \{skill\_based, rule\_based, knowledge\_based\}}\\
\texttt{- initiation: \{self\_initiated, externally\_\\triggered\}}\\
\texttt{- social\_mode: \{solo, synchronous, async, passive\_consumption\}}

\texttt{Recent actions: \{recent\_actions\}}\\
\texttt{Current goals: \{current\_goals\}}\\
\texttt{User Feedback: \{user\_feedback\}}\\
\texttt{User Context: \{user\_context\}}\\
\texttt{Operations: \{operations\}}
\end{quote}

\subsection{Activity Proposal}
\label{app:prompt-activity-propose}

\noindent\textit{Note: The follows, co\_occurs, and overlaps edges between action nodes that this prompt consumes via \{prior\_context\} and \{concurrent\_context\} are constructed deterministically at action-creation time based on timestamp ordering and time-range intersection. They are not inferred by the LLM.}

\begin{quote}
\texttt{Cluster recent ACTIONS into long-lived ACTIVITIES for \{user\_name\}. An ACTIVITY is a {working sphere}---a motive-driven pursuit that organizes multiple actions: stable across runs, broader than a single task but narrower than a life domain, and named with concrete entities (projects, people, organizations, tools).}

\texttt{Inputs:}\\
\texttt{- \{actions\}: current-batch actions, each with \texttt{goal\_hint}, \texttt{conf}, \texttt{decay}, \texttt{dur}, \texttt{object}, \texttt{outcome}, \texttt{domain}, \texttt{community}, \texttt{tension}, \texttt{engage}, \texttt{cog}, \texttt{init}, \texttt{social} fields}\\
\texttt{- \{temporal\_context\}: chronological deltas across the current batch (use as a \emph{weak prior} only — co-occurrence does not imply same working sphere)}\\
\texttt{- \{prior\_context\}: actions from earlier batches reachable via \texttt{follows} edges. Clusterable: their IDs may appear in \texttt{action\_ids}, even if already \texttt{assigned\_to:[<activity\_id>:<label>]} (multi-membership is allowed)}\\
\texttt{- \{concurrent\_context\}: actions linked by \texttt{co\_occurs} or \texttt{overlaps} edges. Also clusterable; may signal coordinated multi-tasking}\\
\texttt{- \{user\_stated\_goals\}, \{system\_goals\}, \\ \{user\_constraints\}, \{user\_context\}: user-provided context and current strivings}

\texttt{Propose NEW activities when 2--3+ actions share a coherent working sphere. Cover {every} current-batch action with at least one candidate.}

\texttt{Per-candidate output (JSON):}\\
\texttt{- description, purpose, reasoning}\\
\texttt{- people, resources, temporal\_pattern}\\
\texttt{- engagement\_profile \{mostly\_sustained,\\ mostly\_fragmented, mixed\}}\\
\texttt{- initiation\_profile \{mostly\_self\_initiated, mostly\_externally\_triggered, mixed\}}\\
\texttt{- identity\_context \{work, personal, creative, social, health\}}\\
\texttt{- action\_ids: ALL actions assigned (current + clusterable context)}\\
\texttt{- action\_valences: parallel array, per action, \{supports, hinders, neutral\}}\\
\texttt{- confidence (1--10): based on evidence breadth, named-entity overlap, and engagement quality}
\end{quote}

\subsection{Activity Reconciliation}
\label{app:prompt-activity-reconcile}

\begin{quote}
\texttt{Reconcile new activity candidates against the existing activity repository. For each candidate, choose ONE decision:}\\
\texttt{- {match}: candidate IS an existing activity}\\
\texttt{- {revise}: same activity, but label is stale---update it}\\
\texttt{- {new}: genuinely different pursuit}\\
\texttt{- {merge}: candidate reveals 2+ existing activities were actually one}

\texttt{Constraints:}\\
\texttt{- [locked] goals: MUST NOT merge, delete, or relabel}\\
\texttt{- [user-edited]: MUST keep exact label}\\
\texttt{- [user-reassigned]: respect current assignment}\\
\texttt{- User annotations outweigh behavioral inference}
\texttt{- NEVER merge two activities if either is [locked]}

\texttt{Candidates: \{candidates\}}\\
\texttt{Existing activities: \{existing\_activities\}} \\
\texttt{\{user\_constraints\}} \\
\texttt{User Context: \{user\_context\}}
\end{quote}

\subsection{Striving Synthesis}
\label{app:prompt-striving-synthesize}

The synthesis prompt is invoked once activities have begun stabilizing after a warmup batch, with a twenty-four-hour ceiling enforced as a fallback.

\begin{quote}
\texttt{Synthesize personal strivings (Emmons, 1986) from observed activities. A striving has three defining properties:}\\
\texttt{1. {Dispositional recurrence}---ongoing, not a one-time objective}\\
\texttt{2. {Active behavioral manifestation}---shows up in what user does}\\
\texttt{3. {Abstraction above actions}---organizes multiple activities}

\texttt{Constraints:}\\
\texttt{- Goals marked [user-edited] or [user-provided]: keep their exact label (you may reassign activities).}\\
\texttt{- Goals marked [locked]: ground truth---do NOT merge, delete, or substantially alter.}\\
\texttt{- User annotations provide privileged \\context---weight higher than inferred evidence.}\\
\texttt{- Activities marked [user-reassigned] should stay with their assigned goal.}\\
\texttt{- Account for every existing goal: if you drop one, list it in \texttt{dropped\_goals} with a reason---never drop without providing a rationale.}

\texttt{Per-striving dimensions:}\\
\texttt{- needs: \{competence, autonomy, relatedness, status, self\_coherence, understanding, order, nurturance, safety, growth, stimulation, purpose\}}\\
\texttt{- orientation: \{approach, avoidance, mixed\}}\\
\texttt{- aspiration\_vs\_obligation: \{ideal, ought, mixed\}}\\
\texttt{- autonomy: \{autonomous, controlled, mixed\}}\\
\texttt{- identity\_link: connection to hoped-for self}\\
\texttt{- feared\_self: what feared self user is avoiding}

\texttt{Activities: \{activities\}}\\
\texttt{Existing strivings: \{existing\_goals\}}\\
\texttt{\{user\_constraints\}}\\
\texttt{\{user\_review\_edits\}}\\
\texttt{User context: \{user\_context\}}
\end{quote}

\subsection{Striving Self-Refine}
\label{app:prompt-striving-refine}

The synthesis output is passed through a self-refine step that critiques and revises the striving set against 21 quality checks organized into four categories:

\begin{quote}
\texttt{Critique the previous synthesis for:}

\texttt{{Standard quality checks:}}\\
\texttt{- Redundancy: should any goals be merged?}\\
\texttt{- Overly broad: is any goal so broad it's meaningless?}\\
\texttt{- Restating an activity: apply the completability test---can it be checked off? If so, it's a project, not a striving.}\\
\texttt{- Too surface-level: push past task categories toward the underlying pursuit.}\\
\texttt{- Missing coverage: are any activities unassigned?}

\texttt{{Phrasing checks:}}\\
\texttt{- Phrasing variety: avoid repeated constructions.}\\
\texttt{- Wish vs.\ pursuit: ``wants to'' describes aspiration, not striving. Rephrase as what user is actively doing.}\\
\texttt{- Behavioral grounding: each goal must describe something the system has evidence of user \emph{doing}.}

\texttt{{Theory-grounded checks:}}\\
\texttt{- Conflict completeness (Emmons): are all key tensions captured?}\\
\texttt{- Avoidance \& ambivalence gap: look for \\ procrastination, defensive activities, or mixed approach/avoidance.}\\
\texttt{- Ideal/ought balance: look for obligation-driven patterns.}\\
\texttt{- Autonomy signal check: look for\\ externally-triggered, low-engagement tasks.}\\
\texttt{- Level appropriateness (Emmons): not so abstract it's a need label, not so concrete it's a project.}
\texttt{{Constraint compliance:}}\\
\texttt{- [locked] goals: MUST NOT merge, delete, or relabel.}\\
\texttt{- [user-edited] labels: keep verbatim.}\\
\texttt{- User annotations outweigh behavioral inference.}

\texttt{Produce a REVISED set of strivings. Keep goals that are good, merge/split/revise those that need it, and ensure every activity is covered.}

\texttt{Previous synthesis: \{previous\_output\}}\\
\texttt{Activities: \{activities\}}\\
\texttt{\{user\_constraints\}}\\
\texttt{\{user\_review\_edits\}}\\
\texttt{User context: \{user\_context\}}
\end{quote}

\subsection{Contextual Integrity Audit}
\label{app:prompt-audit}

\begin{quote}
\texttt{Analyze user input against past interactions for privacy compliance.}

\texttt{Return:}\\
\texttt{- is\_new\_information (bool): contains new information vs.\ past interactions?}\\
\texttt{- data\_type: \{banking\_credentials, \\health\_information, personal\_communications, work\_activity,\\browsing\_activity, general\_activity, none\}}\\
\texttt{- subject: primary subject of disclosed data}\\
\texttt{- recipient: who receives the information}\\
\texttt{- transmit\_data (bool): should data be transmitted based on user's past privacy patterns?}\\
\texttt{- reasoning: brief explanation of decision}

\texttt{User input: \{observation\}}\\
\texttt{Past interactions: \{past\_context\}}
\end{quote}

\section{Participant Background}
\label{app:participants}

Of our 14 participants, 5 identified as women, 7 as men, and 2 as non-binary. Five were aged 18--24, eight were 25--34, and one was 35--44. They represented the following backgrounds: eight PhD students, three postdoctoral researchers, one master's student, one undergraduate, and one startup founder. They spanned a variety of fields including Computer Science (4), AI and Machine Learning (2), HCI (2), Bioengineering (1), Civil Engineering (1), Cognitive Science (1), Economics (1), Physics (1), and Mineral Refinement (1). Twelve participants reported using large language models daily and two reported using them a few times per week. All used Apple Silicon Macs as their primary computer, with all of them using the same machine for both work and personal activities.

\newpage

\section{System Property Graph}
\label{app:property-graph}

\begin{figure}[H]
    \centering
    \includegraphics[width=\linewidth]{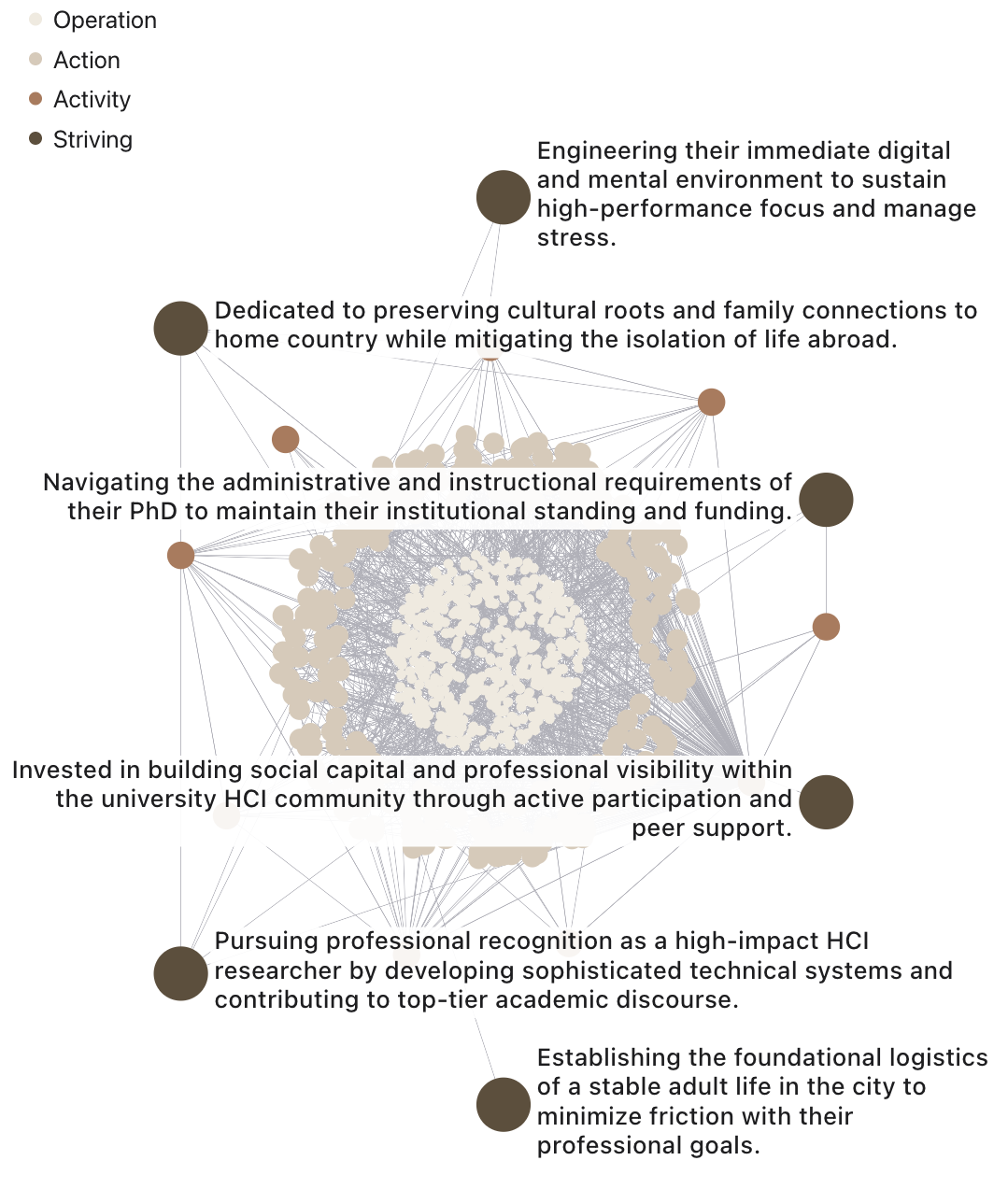}
    \caption{\system{} stores the striving hierarchy as a property graph. Each operation, action, activity, and striving is stored as a node connected by parent-child edges and by the temporal edges (\textit{follows}, \textit{co-occurs}, \textit{overlaps}) the pipeline accumulates between action nodes and between activity nodes as new observations arrive.}
    \label{fig:property_graph}
\end{figure}

% \subsection{Evaluation Models using Linear-Mixed Effects Models.}
\clearpage
\onecolumn

\section{Likert Descriptive Statistics Table Breakdown}
\label{app:likert}

\begin{table}[!htbp]
\centering
\caption{Item-level descriptive statistics across conditions ($N=14$). Ratings on a 7-point Likert scale from $-3$ (Strongly Disagree) to $+3$ (Strongly Agree).}
\label{tab:likert-descriptives}
\small
\begin{tabular}{@{}l l c c c@{\hskip 18pt}l c c@{}}
\toprule
& & \textbf{Tempo} & \textbf{Tempo} & \textbf{Tempo} & & \textbf{Screenshot} & \textbf{Hierarchical} \\
& & $-$\{\textbf{Hierarchy,} & $-$\{\textbf{User} & & & \textbf{View} & \textbf{View} \\
& & \textbf{User Context}\} & \textbf{Context}\} & & & & \\
\cmidrule(lr){3-3} \cmidrule(lr){4-4} \cmidrule(lr){5-5} \cmidrule(lr){7-7} \cmidrule(lr){8-8}
& \textbf{Item} & $M$ ($SD$) & $M$ ($SD$) & $M$ ($SD$) & \textbf{Item} & $M$ ($SD$) & $M$ ($SD$) \\
\midrule
\multicolumn{5}{@{}l}{\textbf{Striving Quality} (per-goal, $\alpha = .86$)} & \multicolumn{3}{l}{\textbf{Editing Experience} (per-participant, $\alpha = .81$)} \\[2pt]
& Accuracy           & 1.37 (1.56) & 1.20 (1.68) & 1.25 (1.76) & Agency           & 1.50 (1.74) & 1.93 (1.21) \\
& Alignment          & 1.38 (1.48) & 1.16 (1.80) & 1.46 (1.50) & Control          & 1.14 (1.51) & 1.79 (1.12) \\
& Abstraction        & 0.79 (1.81) & 1.01 (1.88) & 0.88 (1.93) & Transparency     & 0.00 (1.88) & 2.14 (1.10) \\
& Characteristicness & 1.21 (1.63) & 1.20 (1.73) & 1.35 (1.62) & Evidence utility & 0.79 (2.15) & 2.50 (0.52) \\
\cmidrule(l){2-5}
& \textit{Overall}   & 1.19 (1.64) & 1.14 (1.77) & 1.24 (1.71) & Reflection       & 0.14 (1.66) & 1.00 (1.88) \\
\midrule
\multicolumn{5}{@{}l}{\textbf{Striving Representativeness} (per-set, $\alpha = .79$)} & Confidence       & 0.57 (1.34) & 1.29 (1.38) \\[2pt]
& Captures goals         & 0.93 (1.98) & 1.29 (1.44) & 1.50 (1.70) & Ease             & 1.79 (1.58) & 2.07 (1.14) \\
& Surfaced new connections & $-$0.21 (1.85) & 0.50 (1.61) & 0.36 (1.74) & & & \\
& Reflects motives       & 0.07 (1.98) & 0.79 (1.72) & 0.57 (1.74) & & & \\
\cmidrule(l){2-5} \cmidrule(l){6-8}
& \textit{Overall}       & 0.26 (1.95) & 0.86 (1.59) & 0.81 (1.76) & \textit{Overall} & 0.85 (1.77) & 1.82 (1.30) \\
\bottomrule
\end{tabular}
\end{table}

\section{System Interface}
\label{app:system-fig}

\begin{figure}[!htbp]
    \centering
    \includegraphics[width=\textwidth]{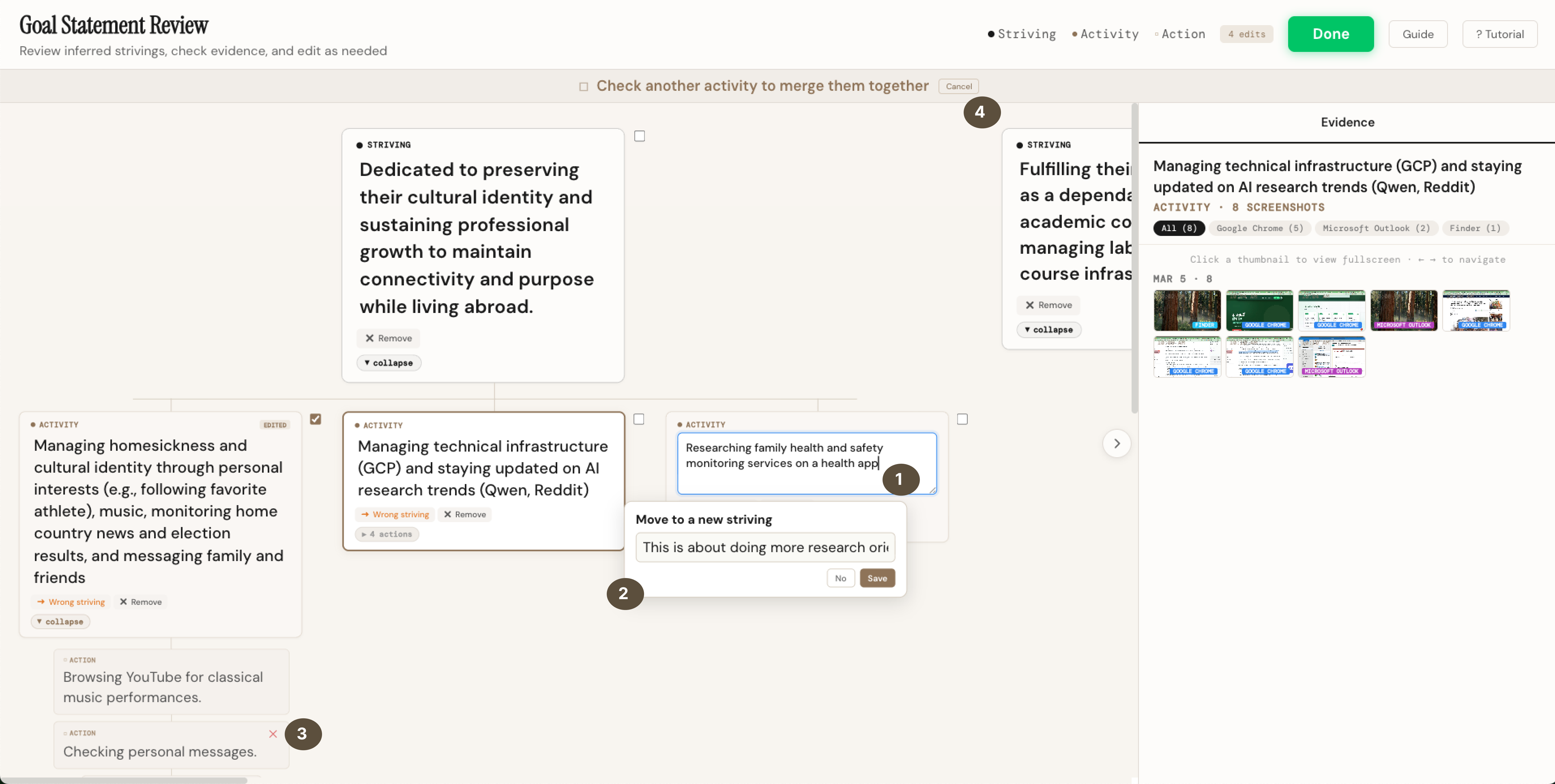}
    \caption{The hierarchical view used in the evaluation of the editing module (\S\ref{sec:eval_edit}). Strivings from the full \system{} are surfaced as a navigable tree of strivings, activities, and actions, with screenshot evidence for the selected node in the right panel. The four numbered callouts illustrate the editing operations (\S\ref{sec:edit}) at the levels where each applies: (1) \textit{inline edit} 
    an activity, (2) \textit{reassign} an activity to a new striving with a free-text justification, (3) \textit{remove} an action, and (4) \textit{merge} two activities or strivings.}
\label{fig:system_tree}
\end{figure}

\begin{figure}[!htbp]    
    \centering
    \includegraphics[width=\textwidth]{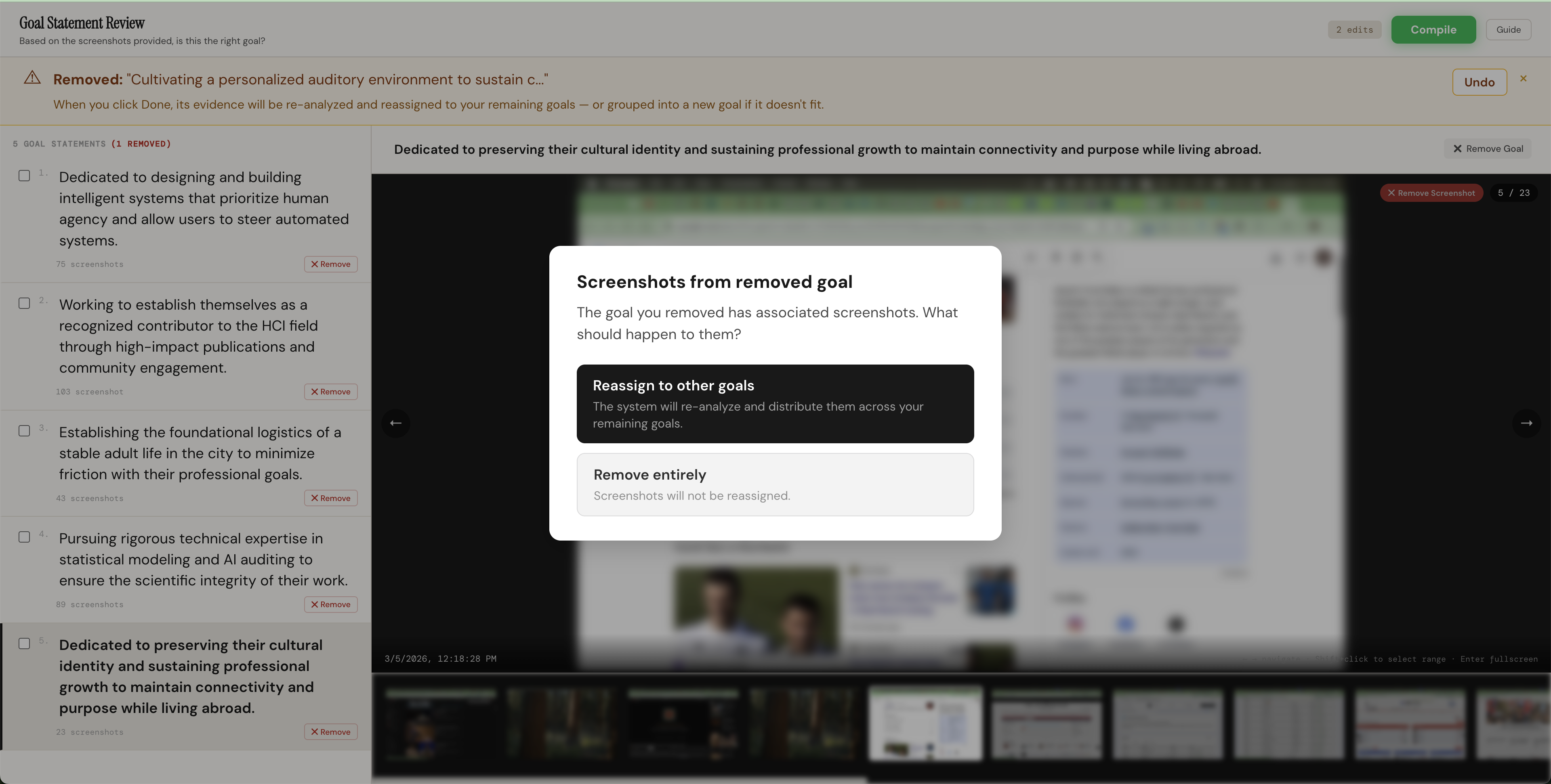}
    \caption{The screenshot view used as the baseline condition in the evaluation of the editing module (\S\ref{sec:eval_edit}), corresponding to the \system{}  $-$ \{Hierarchy, User Context\} ablation (\S\ref{sec:induce_ablations}).  Strivings are listed in the left panel without any intermediate structure,  with the screenshots contributing to each striving displayed in the main viewer.  Users can inline-edit striving text, redistribute or remove individual  screenshots, remove strivings, or merge them. This system screenshot shows what happens when a user removes a striving---a dialog offers the choice to reassign its screenshots to remaining goals or discard them entirely.}
    \label{fig:system_scrsht}
\end{figure}

% \twocolumn

% \newpage 


\end{document}